\documentclass[apj]{emulateapj}
\usepackage{amsmath}
\usepackage{ulem}
\usepackage{xcolor}
\usepackage{natbib}
\usepackage{hyperref}
\hypersetup{
colorlinks=true,
linkcolor=blue,
citecolor=blue}

\newcommand{\ssst}{\scriptscriptstyle}
\newcommand{\E}[1]{\times 10^{#1}}
\newcommand{\etal}{et al.}

\newcommand{\RA}[3]{{#1}^{{\rm h}}{#2}^{{\rm m}}{#3}^{\rm s}}
\newcommand{\decl}[3]{{#1}^{\circ}{#2}'{#3}''}

\newcommand{\s}{\,{\rm s}}      \newcommand{\ps}{\,{\rm s}^{-1}}
    \newcommand{\Msun}{M_{\odot}}
\newcommand{\cm}{\,{\rm cm}}    \newcommand{\km}{\,{\rm km}}
\newcommand{\kms}{$\km\ps$}
 
        \newcommand{\K}{\,{\rm K}}

\newcommand{\um}{\,\mu\rm m}

\newcommand{\nel}{n_{e}}        \newcommand{\NH}{N_{\ssst\rm H}}

\newcommand{\kTc}{kT_{\rm c}} \newcommand{\kTh}{kT_{\rm h}}
 
\newcommand{\tauc}{\tau_{\rm c}} \newcommand{\tauh}{\tau_{\rm h}}
         
\newcommand{\nH}{n_{\ssst\rm H}}        \newcommand{\mH}{m_{\ssst\rm H}}
\newcommand{\nHH}{n({\rm H}_{2})} \newcommand{\NHH}{N({\rm H}_{2})}

\newcommand{\VLSR}{V_{\ssst\rm LSR}}
\newcommand{\Tmb}{T_{\rm mb}}
       
 \newcommand{\Spitzer}{{\sl Spitzer}}

\newcommand{\XMMN}{{\sl XMM-Newton}}
\newcommand{\Chandra}{{\sl Chandra}}

\newcommand{\Fermi}{{\sl Fermi}}
\newcommand{\du}{d_{7.1}} 
    
\newcommand{\Rb}{R_{\rm b}}

\newcommand{\rray}{$\gamma$-ray}

\newcommand{\snr}{Kes~79}
\newcommand{\twCO}{$^{12}$CO}   \newcommand{\thCO}{$^{13}$CO}
\newcommand{\HCOp}{HCO$+$}
\newcommand{\Jotz}{$J=1$--0}    \newcommand{\Jtto}{$J=2$--1}
\newcommand{\Jttt}{$J=3$--2}

\newcommand{\nei}{$nei$}
\newcommand{\pshock}{$pshock$}
\newcommand{\vnei}{$vnei$}
\newcommand{\apec}{$apec$}
\newcommand{\vpshock}{$vpshock$}

\newcommand{\HI}{\ion{H}{1}}

\slugcomment{\sc The Astrophysical Journal, 831:192 (17pp),
published 2016 November 8}

\shortauthors{Zhou et al.}
\shorttitle{Multi-wavelength study of the supernova remnant Kes 79 (G33.6+0.1)}

\begin{document}

\title{Multi-wavelength study of the supernova remnant Kes 79 
(G33.6+0.1): On its supernova properties and expansion into a 
molecular environment}


\author{
  Ping Zhou\altaffilmark{1},
 Yang Chen\altaffilmark{1,2},
 Samar Safi-Harb\altaffilmark{3},
 Xin Zhou\altaffilmark{4,5},
 Ming Sun\altaffilmark{6},
 Zhi-Yu Zhang\altaffilmark{7,8},
 and Gao-Yuan Zhang\altaffilmark{1}
}

\affil{$^1$ School of Astronomy and Space Science, Nanjing University, Nanjing~210023, 
China; \href{mailto:pingzhou@nju.edu.cn}{pingzhou@nju.edu.cn}}
\affil{$^2$ Key Laboratory of Modern Astronomy and Astrophysics,
 Nanjing University, Ministry of Education, China}
\affil{$^3$ Department of Physics and Astronomy, University of
Manitoba, Winnipeg~R3T 2N2, Canada}
\affil{$^4$ Purple Mountain Observatory, CAS, 2 West Beijing 
Road, Nanjing 210008, China}
\affil{$^5$ Key Laboratory of Radio Astronomy, Chinese Academy
of Sciences, Nanjing 210008, China}
\affil{$^6$ Physics Department, University of Alabama in 
Huntsville, Huntsville, AL 35899, USA}
\affil{$^7$ Institute for Astronomy, University of Edinburgh, 
Royal Observatory, Blackford Hill, Edinburgh EH9 3HJ, UK}
\affil{$^8$ ESO, Karl Schwarzschild Strasse 2, D-85748 Garching, 
Munich, Germany}

\begin{abstract}
\snr\ (G33.6+0.1) is an aspherical thermal composite supernova remnant (SNR) observed across
the electromagnetic spectrum and showing an unusual highly structured
morphology, in addition to harboring a central compact object (CCO).
Using the CO \Jotz, \Jtto, and \Jttt\ data, we provide the first direct
evidence and new morphological evidence to support the physical interaction
between the SNR and the molecular cloud in the local standard of rest 
velocity $\sim 105$~\kms.
We revisit the 380~ks \XMMN\ observations and perform a dedicated
spatially resolved X-ray spectroscopic study with careful background
subtraction.
The overall X-ray-emitting gas is characterized by an under-ionized
($\tauc \sim 6\E{11}~\cm^{-3}$) cool ($\kTc\approx 0.20$~keV) plasma 
with solar abundances, plus an under-ionized ($\tauh\sim 8\E{10}~\cm^{-3}$) 
hot ($\kTh\approx 0.80$~keV) plasma 
with elevated Ne, Mg, Si, S and Ar abundances.
The X-ray filaments, spatially correlated with the 24~$\um$ IR
filaments, are suggested to be due to the SNR shock interaction 
with dense gas, while the halo forms from SNR breaking out into
a tenuous medium.
Kes~79 appears to have a double-hemisphere morphology viewed
along the symmetric axis.
Projection effect can explain the multiple-shell structures and 
the thermal composite morphology.
The high-velocity, hot ($\kTh\sim 1.4$--1.6~keV) ejecta patch
with high metal abundances, together with the 
non-uniform metal distribution across the SNR, indicates an asymmetric 
SN explosion of \snr.
We refine the Sedov age to 4.4--6.7~kyr and the mean shock velocity
to 730~\kms.
Our multi-wavelength study suggests a progenitor mass of $\sim 15$--20 
solar masses for the core-collapse explosion that formed \snr\
and its CCO, PSR~J1852+0040.
\end{abstract} 


\section{Introduction} \label{S:intro}

Core-collapse supernova remnants (SNRs) are more or less aspherical
in their morphology (Lopez \etal\ 2011).  
The asymmetries could be caused by external shaping from 
non-uniform ambient medium (Tenorio-Tagle \etal\ 1985), 
the dense slow winds of progenitor stars (Blondin \etal\ 1996), 
the runaway progenitors (Meyer \etal\ 2015),
and the Galactic magnetic field (Gaensler 1998; West \etal\ 2015).
Intrinsic asymmetries of the explosion can also impact on the 
morphologies of SNRs, with increasing evidences 
provided by studying the distribution and physical states of the ejecta.
The historical SNR Cas~A shows fast moving ejecta knots outside the 
main shell (e.g.\ , Fesen \& Gunderson 1996) and non-uniform distribution
of heavy elements (e.g. Hwang \etal\ 2000; $^{44}$Ti recently
reported by Grefenstette \etal\ 2014).
High-velocity ejecta ``shrapnels'' have been discovered in the evolved
SNR Vela (Aschenbach \etal\ 1995).
The accumulating observations of asymmetric SNRs challenge the standard
spherical pictures of SN explosion and SNR evolution.
In light of this, the environmental and spatially resolved study 
of asphercial SNRs becomes more and more important.

\snr\ (a.k.a.\ G33.6+0.1) is a Galactic SNR with a round 
western boundary and deformed eastern boundary in radio band
(Frail \& Clifton 1989).
The radio morphology is characterized by multiple concentric shells
or filaments (Velusamy \etal\ 1991).
An early ROSAT X-ray observation showed that most of the diffuse X-ray emission is
from a bright inner region and some faint X-ray emission is extended to
the outer region (Seward \& Velusamy  1995).
The 30~ks \Chandra\ observation revealed rich structures, such as
many filaments  and a ``protrusion,'' and a constant temperature
(0.7~keV) across the SNR (Sun \etal\ 2004; hereafter S04).
The spectral results of the global SNR were next supported with 
the spectral study using two epochs of \XMMN\ observations 
(Giacani \etal\ 2009).
Using \XMMN\ observations spanning 2004 and 2009,
the spatially resolved studies provided further information on
the hot gas, where two thermal components are required to explain the
observed spectra (Auchettl \etal\ 2014, hereafter A14).
\snr\ hosts a central compact object (CCO) PSR~J1852+0040 (Seward \etal\ 
2003), which was discovered as a 105~ms X-ray pulsar 
(Gotthelf \etal\ 2005) with a weak magnetic field (Halpern \etal\ 2007; 
Halpern \& Gotthelf 2010).
In the south, an 11.56~s low-B magnetar, 3XMM~J185246.6+003317, was
found at a similar distance to \snr\ (Zhou \etal\ 2014; Rea \etal\ 2014).

\snr\ is classified as a thermal composite (or mixed-morphology) SNR 
presenting a centrally filled morphology in X-rays and shell-like 
in the radio band (Rho \& Petre 1998).
Thermal composite SNRs generally display good correlation with \HI\ or molecular
clouds (MCs; Rho \& Petre 1998; Zhang \etal\ 2015) and \rray\ emission 
(check the SNR
catalog\footnote{\url{http://www.physics.umanitoba.ca/snr/SNRcat/}} in
Ferrand \& Safi-Harb 2012), and are believed to be the best
targets to study hadronic cosmic rays.
Green \& Dewdney  (1992) performed \twCO~\Jotz\ and \HCOp~\Jotz\ observations
toward \snr\ and found a morphological coincidence of the SNR with 
the MCs in the east and southeast at the local standard of rest
(LSR) velocity ($\VLSR$) of 105~\kms.  
At this velocity, broad 1667~MHz OH absorption (Green 1989) and 
\HI\ absorption were detected (Frail \& Clifton 1989).
Green \etal\ (1997) reported the detection of 1720~MHz OH line emission
toward \snr, although no OH maser was found.
Hence, \snr\ is very likely interacting with MCs at
around 105~\kms, but direct physical evidence is still lacking.
The LSR velocity corresponds to a kinetic distance of 7.1~kpc 
according to the rotation curve of the Galaxy (Frail \& Clifton 1989; Case \&
Bhattacharya 1998).
GeV \rray\ emission is also detected with \Fermi\ east of \snr\ 
at a significance of $\sim7\sigma$, where bright CO emission is present (A14).

In order to study the origin of asymmetries and thermal composite
morphology of \snr\ and to find physical evidence for the SNR--MC 
interaction, we performed new multi-transition CO observations (see
also Chen \etal\ 2014) and revisited the \XMMN\ data.
This paper is organized as follows. 
In Section~\ref{S:obs}, we describe the multi-wavelength observations
and data reduction.
Our results are shown in Section~\ref{S:result} and
the discussion is presented in Section~\ref{S:discussion}. 
A summary is given in Section~\ref{S:summary}.

\section{Observation} \label{S:obs}
\subsection{CO Observations} \label{S:coobs}
The observations of \twCO\ \Jotz\ (at 115.271~GHz)
and \thCO\ \Jotz\ (at 110.201~GHz) were taken during 
2011 May with the 13.7 m millimeter-wavelength telescope of the Purple 
Mountain Observatory at Delingha (PMOD), China.
The new Superconducting Spectroscopic Array Receiver with $3\times 3$ beams 
was used as the front end (Shan \etal\ 2012).
The \twCO\ \Jotz\ and \thCO\ \Jotz\ lines were configured at at the upper and 
lower sidebands, respectively, and observed simultaneously.
The Fast Fourier Transform Spectrometers with 1 GHz bandwidth and 
16,384 channels were used as the back ends, providing a velocity 
resolution of 0.16 \kms\ for \twCO\ \Jotz\ and 0.17 \kms\ for \thCO\ \Jotz.
The full width at half maximum (FWHM) of the main beam was $56''$, the 
main beam efficiency was 0.48 in the observation epoch, and 
the pointing accuracy was better than $4''$.
We mapped a $30'\times 30'$ region centered at ($\RA{18}{52}{40}$, 
$\decl{+00}{38}{42}$, J2000) in on-the-fly observing mode.
The data were then converted to \twCO\ \Jotz\ and \thCO\ 
\Jotz\ main-beam temperature ($T_{\rm mb}$) cubes with a grid spacing 
of $0\farcm{5}$ and a velocity resolution of 0.5 \kms.
The corresponding mean rms of the spectra in each pixel is 
0.41~\K\ and 0.26~\K\ per channel, respectively.

The \twCO\ \Jtto\ (at 230.538~GHz ) observation of \snr\ was carried
out in 2010 January with K\"olner Observatory for Submillimeter 
Astronomy (KOSMA; now renamed CCOSMA).
We mapped an $18'\times 16'$ area centered at ($\RA{18}{52}{35}$,
$\decl{+00}{39}{12}$, J2000) with a grid spacing of $1'$ using
a superconductor--insulator--superconductor receiver and a 
medium-resolution acousto-optical spectrometer.
The FWHM of the main beam was $130''$ and the velocity resolution
was $0.21$ \kms\ at 230~GHz.
The main beam efficiency was $68\%$.

We retrieve the \twCO~\Jttt\ (at 345.796~GHz) data of \snr\ from the 
CO High Resolution Survey (Dempsey \etal\ 2013) observed with Heterodyne 
Array Receiver Program (HARP) on board the James Clerk Maxwell Telescope (JCMT). 
The first release of the survey covers $|b|\le 0\fdg{5}$ between
$10\fdg{25}<l<17\fdg{5}$, and $|b|\le 0\fdg{25}$ between $17\fdg{5}
<l<50\fdg{25}$.
The FWHM of the main beam at the frequency was $14''$, and the 
main beam efficiency was 0.63.
We use the rebinned data, which has a pixel size of 6$''$ and a velocity 
resolution 1~\kms.

The \Jotz\ and \Jtto\  data were reduced with the GILDAS/CLASS 
package.\footnote{\url{http://www.iram.fr/IRAMFR/GILDAS/}}
For the data analysis and visualization, we made use of the  
Miriad (Sault \etal\ 1995), 
KARMA (Gooch 1996) 
and IDL software packages.

\subsection{The {\rm XMM-Newton} and {\rm Spitzer} Data}
There are 21 archival \XMMN\ observations toward \snr, which were taken in
2004 (PI: F. Seward), 2006-2007 (PI: E. Gotthelf), and 2008-2009 (PI:
J. Halpern).  We excluded five observations (Obs.\ ID:
550671401,550671501, 55061601, 55061701, and 55062001) since they
were performed with
filter cal-closed (affected by radiation) and observation
550671101 suffered from flares for most of the observation time.
We use EPIC-MOS data which covered the whole SNR in full frame mode 
with medium filter, while EPIC-pn 
data are not used here since the pn detector only covered the inner part 
the remnant in small window mode.

The EPIC-MOS data were reduced using the Science Analysis
System software (SAS\footnote{\url{http://xmm.esac.esa.int/sas/}}). We used
the \XMMN\ Extended Source Analysis Software, XMM-ESAS, in the SAS
package to filter out the time intervals with soft proton
contaminations, detect point sources in the field of view, and create
particle background images.  After removing time intervals with
heavy proton flarings, the total effective exposures are 376 ks and
380 ks for MOS1 and MOS2, respectively.  Detailed information,
including the observation ID, observing date, and net exposure of the observations,
is summarized in Table~\ref{T:xmmobs}.  

We retrieve \Spitzer\ $24 \um$ post-basic calibrated data from the
\Spitzer\ archive. The mid-IR observation was performed as a 24 Micron
Survey of the Inner Galactic Disk Program (PID: 20597; PI: S. Carey).

\section{Results} \label{S:result}

\begin{center}
\begin{deluxetable*}{lcccccc}
\tabletypesize{\footnotesize}
\tablewidth{\textwidth}
\tablecaption{Observational Information of the \XMMN\ Data}
\tablewidth{0pt}
\tablehead{
\colhead{Obs. ID}  & Obs. Date &\multicolumn{2}{c}{Pointing (J2000)}  &
\multicolumn{2}{c}{Exposure$^a$ (ks)} \\ \cline{3-4} \cline{5-6}
 & & R.A. & Decl. & MOS1 & MOS2
}
\startdata
0204970201 & 2004 Oct 17 & 18:52:43.87 & 00:39:07.0 & 30.2 & 29.8  \\
0204970301 & 2004 Oct 23 & 18:52:44.07 & 00:39:08.5 & 27.6 & 27.9  \\
0400390201 & 2006 Oct 08 & 18:52:43.03 & 00:39:00.0 & 25.1 & 24.9  \\
0400390301 & 2007 Mar 20 & 18:52:34.77 & 00:41:47.3 & 30.3 & 30.7  \\
0550670201 & 2008 Sep 19 & 18:52:42.42 & 00:38:50.8 & 20.4 & 20.4  \\
0550670301 & 2008 Sep 21 & 18:52:42.38 & 00:38:50.1 & 29.3 & 29.2  \\
0550670401 & 2008 Sep 23 & 18:52:42.46 & 00:38:53.6 & 34.7 & 34.2  \\
0550670501 & 2008 Sep 29 & 18:52:42.65 & 00:38:53.5 & 31.8 & 32.4  \\
0550670601 & 2008 Oct 10 & 18:52:43.03 & 00:39:04.2 & 30.8 & 30.5  \\
0550670901 & 2009 Mar 17 & 18:52:34.94 & 00:41:45.8 & 22.0 & 23.3  \\
0550671001 & 2009 Mar 16 & 18:52:34.88 & 00:41:48.1 & 19.1 & 20.0  \\
0550671201 & 2009 Mar 23 & 18:52:34.62 & 00:41:44.5 & 15.2 & 15.7  \\
0550671301 & 2009 Apr 04 & 18:52:34.11 & 00:41:40.4 & 19.1 & 20.0  \\
0550671801 & 2009 Apr 22 & 18:52:33.49 & 00:41:29.6 & 27.1 & 27.1  \\
0550671901 & 2009 Apr 10 & 18:52:33.84 & 00:41:35.7 & 13.0 & 14.4  \\ \hline
0550671101$^b$ & 2009 Mar 25 & 18:52:34.39 & 00:41:41.0 & $\le 4.8$ & $\le 6.1$  \\
0550671401$^c$ & 2009 Mar 17 & 18:52:34.95 & 00:41:45.7 & 0 & 0  \\
0550671501$^c$ & 2009 Mar 25 & 18:52:34.36 & 00:41:42.6 & 0 & 0 \\
0550671601$^c$ & 2009 Mar 23 & 18:52:34.61 & 00:41:44.3 & 0 & 0  \\
0550671701$^c$ & 2009 Apr 04 & 18:52:34.11 & 00:41:40.2 & 0 & 0  \\
0550672001$^c$ & 2009 Apr 10 & 18:52:33.85 & 00:41:35.6 & 0 & 0 
\enddata
\label{T:xmmobs}
\tablecomments{
  \tablenotetext{a}{\phantom{0}Effective exposure time.}
  \tablenotetext{b}{\phantom{0}The observation was not used because
it suffers soft proton contamination for most of the observation time.}
  \tablenotetext{c}{\phantom{0}The observation was not used because
it was affected by radiation and the filter was in cal-closed mode.}
}
\end{deluxetable*}
\end{center}

\subsection{MCs at $\sim 95$--115~\kms}  \label{S:mol}

\snr\ has been suggested to expand into the denser ISM in the east,
since the eastern radio shells are deformed and has a prominent 
protrusion in the northeast (Velusamy \etal\ 1991).
Previous CO \Jotz\ observations showed that the SNR morphology is 
correlated with the eastern molecular gas at
$\VLSR\sim 105~\km\ps$ (Green \& Dewdney 1992).
In this paper, we present the direct kinematic evidence 
with multiple transitions of CO.

\begin{figure*}[tbh!]
\vspace{-0.6in}
\centering
\includegraphics[angle=0, width=0.97\textwidth]{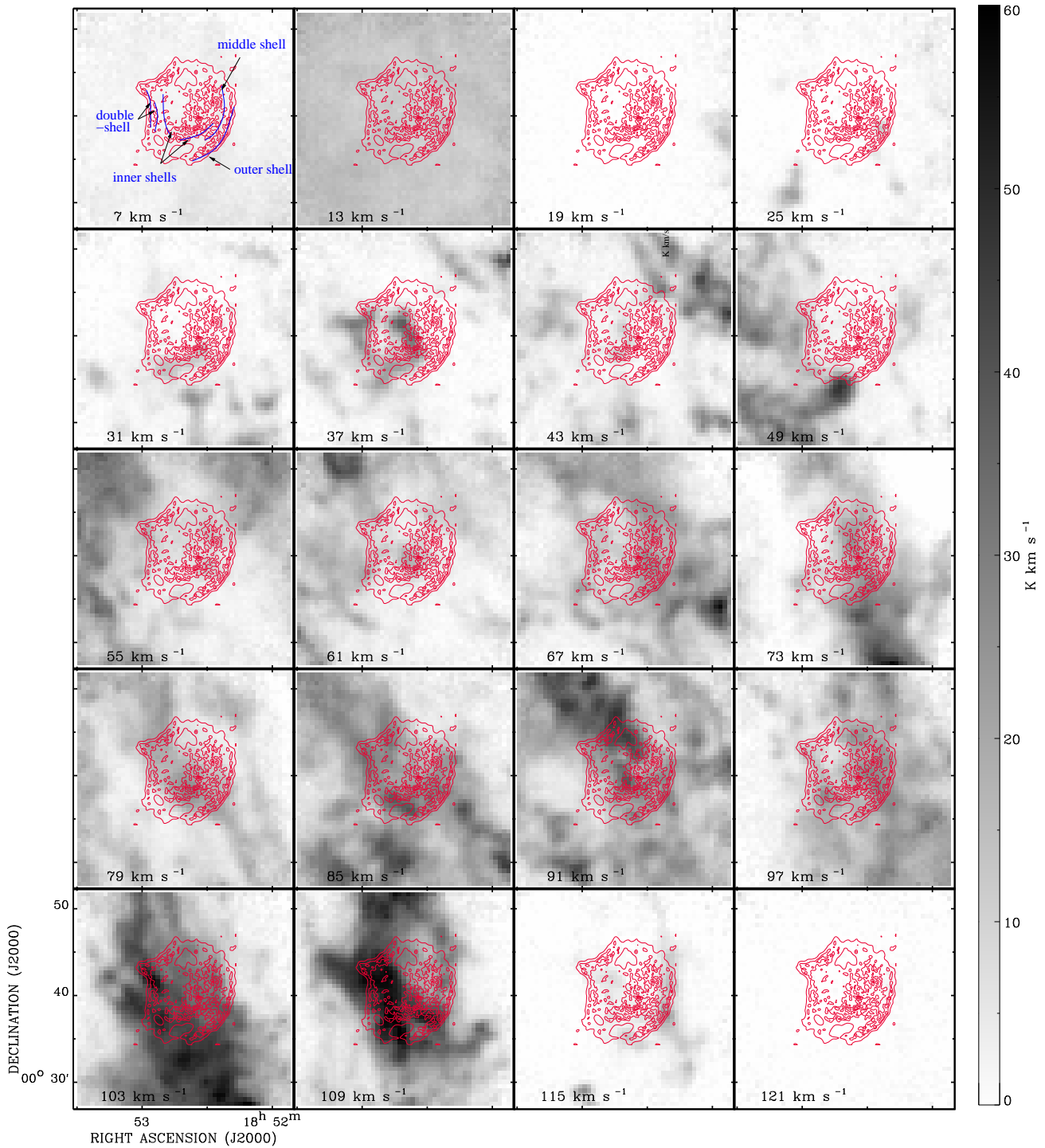}
\vspace{-0.4in}
\caption{
\twCO\ \Jotz\ channel maps in the velocity range 4--124~\kms,
overlaid with contours of the Very Large Array 20 cm  continuum emission
(Velusamy et al. 1991).
The color bar on the right side shows the intensity scale of
the channel maps (3$\sigma$--60~$\K$; here $\sigma$ is the
root mean square (rms) of the spectrum at each pixel).
In the upper-left panel, we define several radio shells which
are used for multi-wavelength comparison.
}
\label{f:12co10_grid}
\end{figure*}

We first search for morphological coincidences between the CO
emission and the SNR shells, especially for the deformed eastern 
and northeastern shells, which might be shaped by dense ambient medium.
Figure~\ref{f:12co10_grid} displays the channel maps of PMOD \twCO~\Jotz\
in the velocity ranges of $4$--124~\kms\ with a velocity step of 6~\kms.
Each channel shows the velocity-integrated $\Tmb$ in a $24'\times 24'$ mapping
area and the velocity labeled in the channel image indicates the central velocity.
In the upper-left panel, we define several radio shells
(also referring to the definition in S04),
which will be used for multi-wavelength comparison in the following study.
Located on the inner Galactic plane, the sky region of \snr\ is rich in 
molecular gas in its line of sight.
There is bright \twCO\ emission in the eastern and northeastern shells
in the velocity range $100$--112~\kms, as noted by Green \& Dewdney (1992), 
while the CO emission fades out to the northwest.
We also notice that a molecular shell at $\sim 115$~\kms\
spatially matches the SNR's middle and outer radio shells in the
west, and the inner radio shell in the east.
At $\VLSR\sim 35 \km \ps$ and $\sim 49~\km\ps$, the MCs are
spatially close to the eastern double shell of the remnant as well;
however, the gas does not match the deformed northeastern shell.
The CO emission at other velocities is either outside the SNR or 
revealing complicated morphology inside the remnant. 
The open \HI\ shell detected in the north and west at $\sim 95 \km\ps$
(Giacani \etal\ 2009), however, seems to not correspond to shell 
structure of CO emission.
From both the morphology and the spectral lines, we cannot find a signal
for MC--SNR interaction below 95 \kms.

\begin{figure*}[tbh!]
\vspace{-0.6in}
\centering
\includegraphics[angle=0, width=0.97\textwidth]{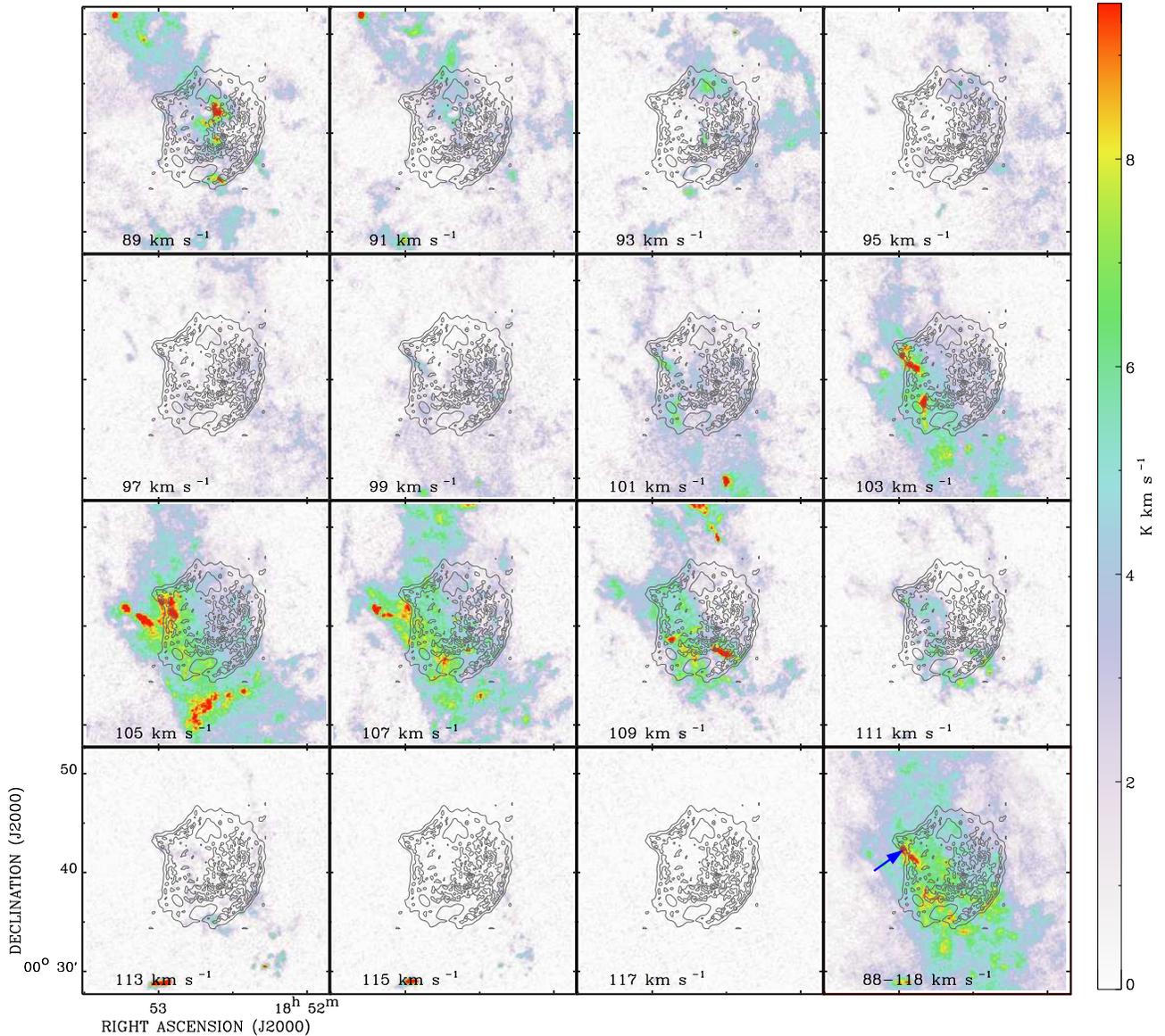}
\vspace{-0.3in}
\caption{
\twCO\ \Jttt\ intensity channel maps in the velocity range 
99.5--114.5~\kms, overlaid with contours of the Very Large Array
20 cm continuum.
The color bar on the right side shows the intensity scale of
the channel maps (3$\sigma$--9.5~$\K$; here $\sigma$ is the
rms of the spectrum at each pixel).
The bottom-right panel shows the velocity-integrated intensity.
The arrow denotes a clump showing a broad \twCO\ \Jttt\
line profile (see Figure~\ref{f:2gaussfit}).
}
\label{f:12co32_grid}
\end{figure*}

The JCMT high-resolution observations in the \twCO~\Jttt\ line (with
beamwidth of $14''$ and pixel size of $6''$) provide more details on the 
gas morphology and can be compared with the VLA 20 cm continuum 
data with a synthesis beam size of $14''\times 7''$.  
Since the \twCO~\Jttt\ transition has a relatively high critical density
of $4\E{4} (T/33~\K)^{-0.5} \cm^{-3}$ and a high upper state 
temperature of 33~K (Sch\"oier \etal\ 2005),
it can trace dense and warm gas better than \twCO~\Jotz,
and thus help to reveal shocked gas.
Figure~\ref{f:12co32_grid} shows HARP \twCO~\Jttt\ channel maps 
in the velocity interval 88--118~\kms, overlaid with the 20~cm 
radio continuum contours.
The velocity step of the channels is 2~\kms\ and the maps 
cover a $24'\times 24'$ area.
The brightest \twCO~\Jttt\ emission is shown near the northeast radio 
protrusion, and along the eastern double shell.
Notably, at $\VLSR\sim 105~\km\ps$, two thin, clumpy \twCO~\Jttt\ filaments
mainly align with the boundary of the eastern double shell,
indicating a relation between the dense and warm molecular filaments
and the eastern double-shell.
We also found that the \twCO~\Jttt\ emission shifts from the 
northeast to the west with increasing velocity (up to 115~\kms).

\begin{figure}
\centering
\includegraphics[angle=0, width=0.47\textwidth]{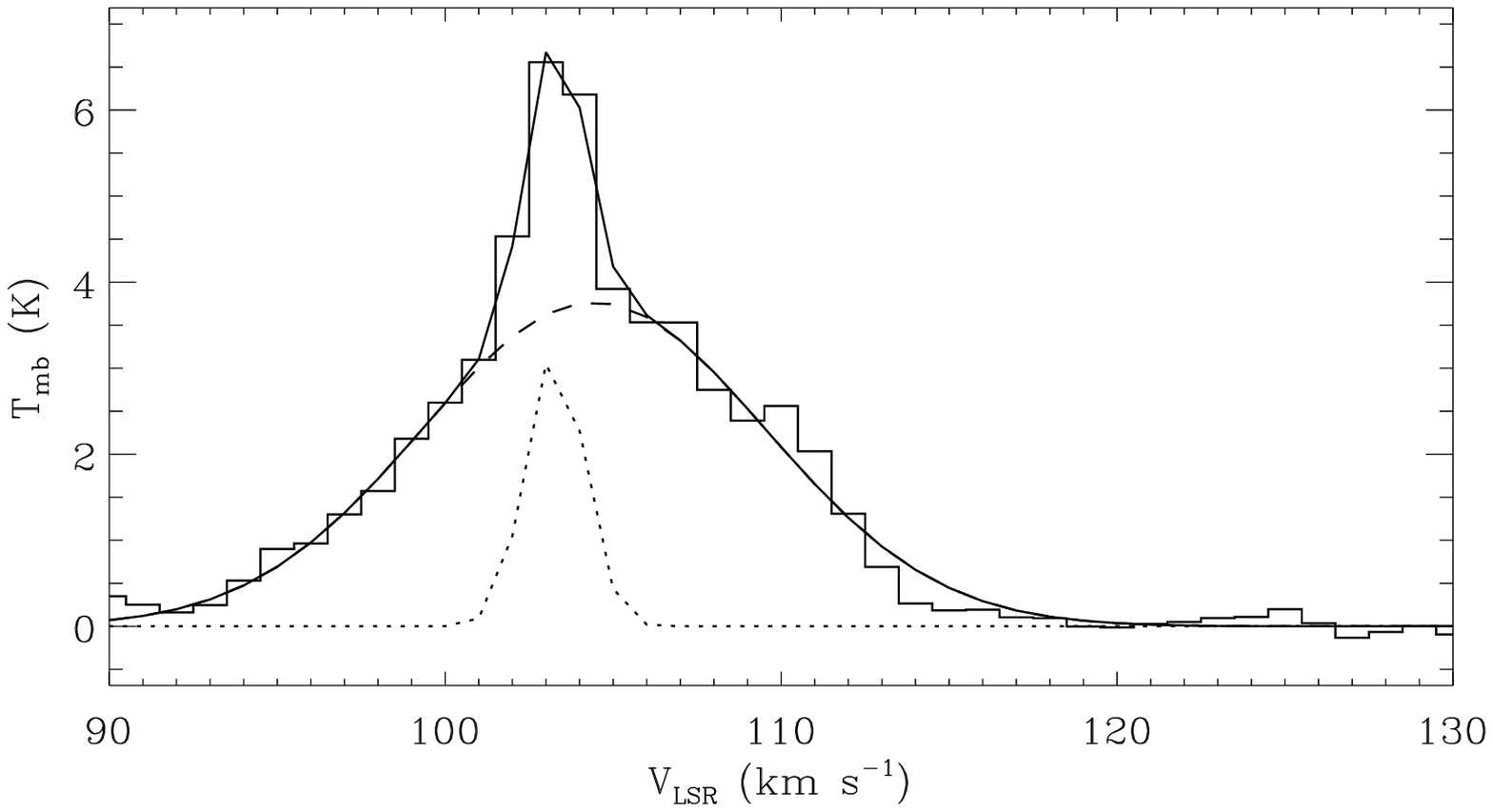}
\caption{
The \twCO~\Jttt\ spectrum of the clump at around the radio protrusion
(denoted by the arrow in Fig~\ref{f:12co32_grid}) fit with a two-Gaussian
model (solid line). 
The narrow component with $\Delta v\approx 2~\km\ps$ and the
broad component with $\Delta v\approx 12~\km\ps$ are shown 
with dotted line and dashed line, respectively.
}
\label{f:2gaussfit}
\end{figure}

Broad molecular line broadening or asymmetric profile is an
evidence of shock perturbation of molecular gas (e.g., Denoyer
1979; Chen \etal\ 2014).
A broad \twCO~\Jttt\ line is found in the bright clump at the southern
boundary of the radio protrusion ($\RA{18}{52}{58}58$, 
$\decl{00}{42}{29}95$, J2000; with size of $\sim 30''$ and denoted by 
an arrow in Figure~\ref{f:12co32_grid}).
As shown in Figure~\ref{f:2gaussfit}, the spectrum consists of
a sharp peak at 103~\kms\ and a broad wing spanning a velocity 
range over 20~\kms, which cannot be explained with a single excitation
component.
We find that the profile can be well described by two Gaussian components:
a narrow line with line width (FWHM) $\Delta v\approx 2$~\kms\ 
at $\VLSR\approx 103.3$~\kms;
and a broad line with $\Delta v\approx 12$~\kms\ at $\VLSR\approx
104.4$~\kms. 
The narrow line corresponds to the dense, quiescent gas, while the 
broad component is probably emitted by the shocked gas.
The broad \twCO~\Jttt\ line detected at the protrusion position,
together with the morphological agreement between the two molecular 
filaments in the east and the radio double shell, supports that the 
systemic velocity of \snr\ is at $\sim 105$~\kms.

\begin{figure*}[tbh!]
\centering
\includegraphics[angle=0, width=\textwidth]{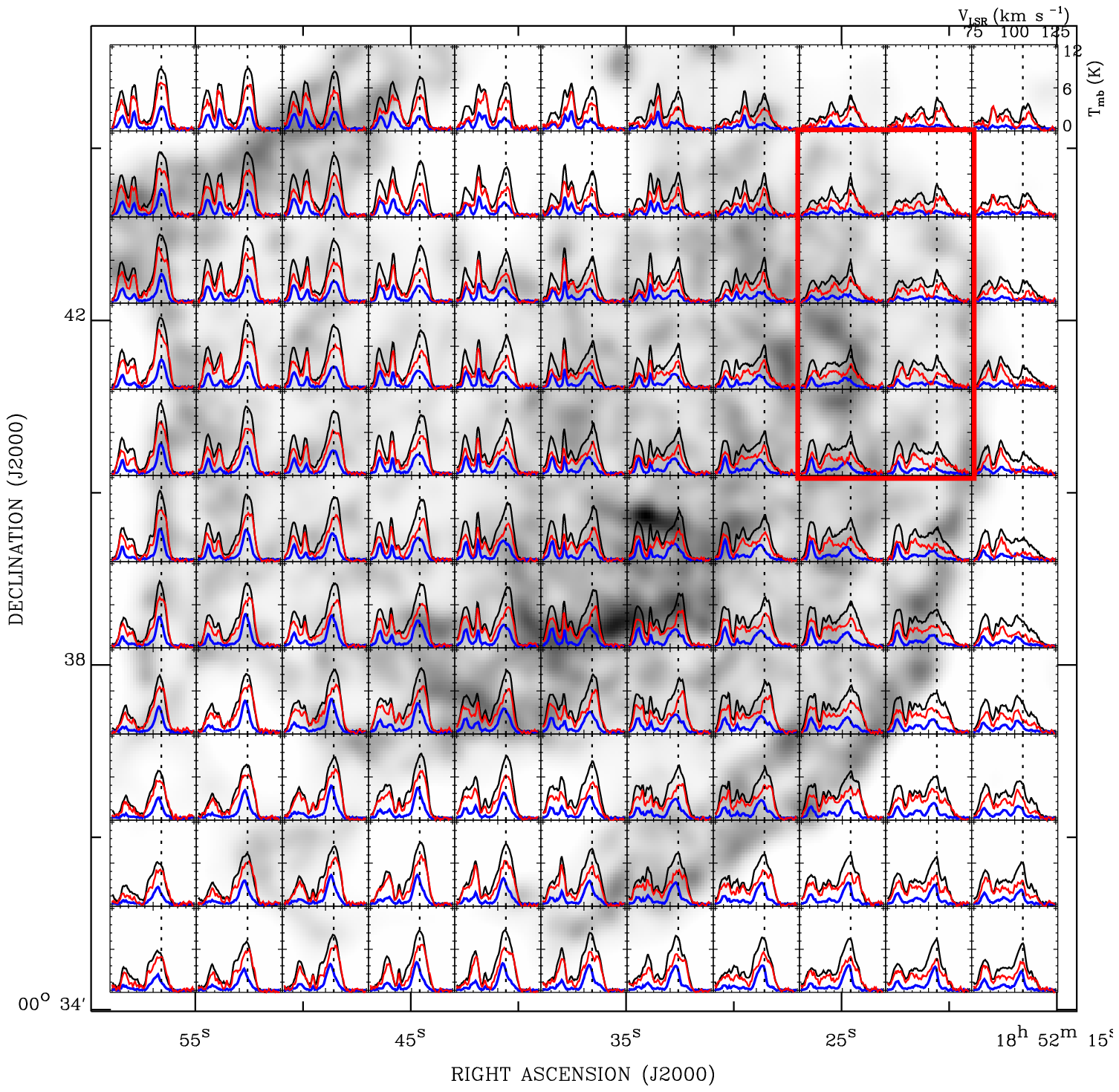}
\caption{\twCO\ \Jotz\ (black), \twCO\ \Jtto\ (red) and \thCO\ 
(blue) spectra in the 75--125~\kms\ velocity range, superposed on
the VLA 20~cm image. 
The scales of $\VLSR$ and $\Tmb$ are labeled in the 
upper-right corner.
The \Jotz\ data cubes were convolved
with a Gaussian kernel to match the beam size of the \twCO~\Jtto\ 
($130''$). Here the pixel size is $60''$. The vertical dashed
lines denote the velocity 105~\kms. 
The red rectangle highlights spectra showing strongly asymmetric
features extended to $\sim 120~\km\ps$.
}
\label{f:linegrid}
\end{figure*}

\begin{figure*}[tbh!]
\centering
\includegraphics[angle=0, width=0.8\textwidth]{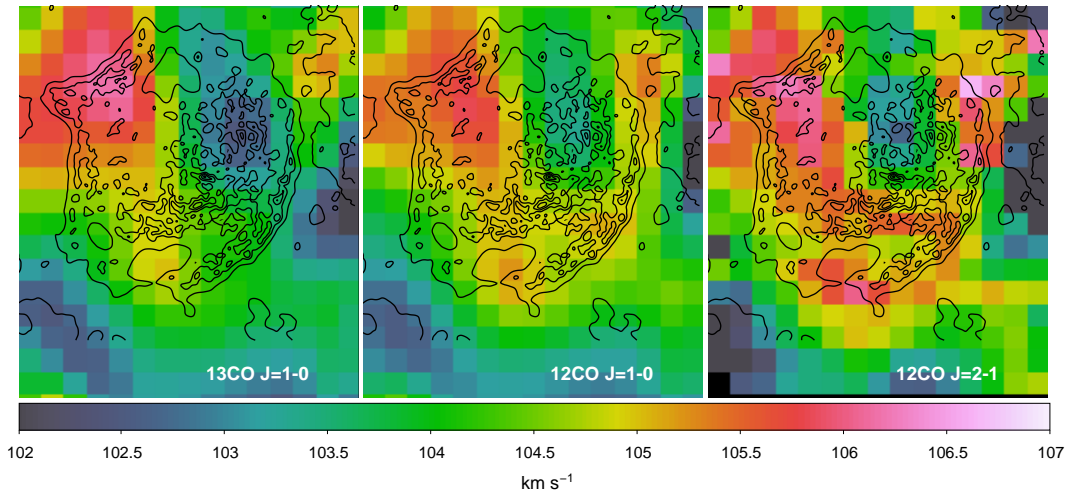}
\caption{
Velocity field (first moment) maps of \thCO\ \Jotz\ (left),
\twCO\ \Jotz\ (middle) and \twCO\ \Jtto\ (right) in the velocity
range 95--115~\kms.
The \Jotz\ data cubes were convolved with a Gaussian kernel to match
the beam size of the \twCO~\Jtto\ ($130''$). Here the pixel size is $60''$.
}
\label{f:moment1}
\end{figure*}

We subsequently search for shocked molecular gas over the entire SNR.
Figure~\ref{f:linegrid} shows a grid of line profiles of \twCO~\Jotz, 
\thCO~\Jotz, \twCO~\Jtto\ in the 75--125~\kms\ velocity range across 
the whole SNR.
The radio image is displayed as a background.
Due to a low abundance of \thCO\ in the interstellar molecular gas
([$^{13}$CO]/[H$_2] \sim 2\E{-6}$; Dickman 1978), the \thCO\ emission is 
normally optically thin and traces the quiescent gas.
The molecular gas shocked by the SNR may show asymmetric or broad 
\twCO\ lines deviating from the line profiles of \thCO.
The \thCO\ emission associated with the SNR peaks at $\VLSR\sim
105~\km\ps$ across the remnant.
In the western hemisphere of the remnant, the \twCO\ lines shift to 
higher velocity (red-shift) relative to the \thCO\ line peaks.
The \twCO\ line profiles in the northeast of the remnant 
(in the red rectangle in Figure~\ref{f:linegrid}) show strongly
asymmetric features and appear to extend to 120~\kms\ 
Figure~\ref{f:moment1} reveals the distribution intensity weighted 
mean velocity (first moment; velocity field) of \thCO\ and \twCO\ lines.
The \twCO\ emission, especially \twCO~\Jtto, has a higher mean velocity 
than the systemic velocity given by the \thCO\ emission in the west.
Notably, the high-$\VLSR$ \twCO\ emission delineates
the outer, middle, and the eastern inner radio shells,
suggesting this high-$\VLSR$ emission 
is excited by the shock at the SNR shells.

\subsection{Multi-wavelength Imaging}

\subsubsection{Filaments Revealed in X-Ray and Other Bands} \label{S:image}

\begin{figure*}[tbh!]
\centering
\includegraphics[angle=0, width=0.9\textwidth]{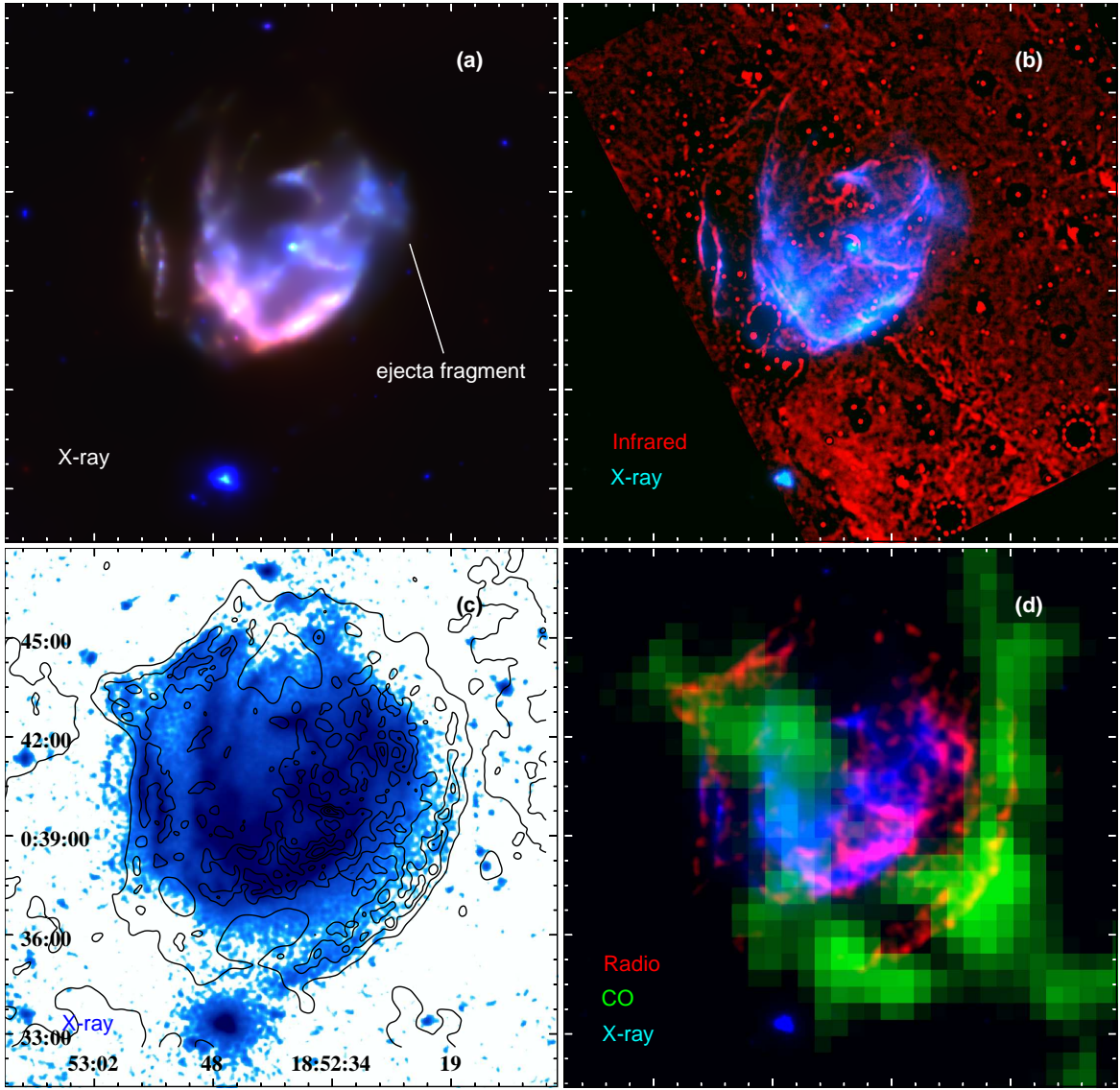}
\caption{
a) Exposure-corrected and adaptively smoothed EPIC-MOS X-ray image of 
\snr.  Red: 0.3-1.2 keV; green: 1.2--2.0 keV; blue: 2.0-7.0 keV.
The blue patch in the northwest labeled with ``ejecta fragment'' is
discussed in Section~\ref{S:fragment}.
b) A composite image of \snr\ showing the spatial correspondence 
between the unsharp mask 24~$\mu$m IR image (red) and the 0.3--7.0
keV X-rays (blue).
c) 0.3--7.0 keV X-ray image shown on a histogram equalization scale
and overlaid with solid, black contours of VLA 20 cm radio continuum
emission.
d) A tri-color image compares the spatial distribution of the
high-$\VLSR$ (110--120~\kms) \twCO~\Jotz\ emission (green) 
to those of the radio (red) and 0.3--7.0~keV X-ray emission (blue).
}
\label{f:rgb}
\end{figure*}

We merged the data of the 15 EPIC-MOS observations and show a tri-color
image of \snr\ in Figure~\ref{f:rgb}(a), in which the 
X-ray emission in the soft (0.3--1.2 keV), medium (1.2--2.0 keV),
and hard (2.0--7.0 keV) bands is colored red, green, and blue,
respectively.
The energy bands are chosen to obtain similar counts in these images.
Each of the three band images was exposure-corrected and adaptively
smoothed with a Gaussian kernel to achieve a minimum significance of
3 and a maximum significance of 4 using the ``csmooth'' command 
in CIAO.\footnote{http://cxc.harvard.edu/ciao}
The X-ray emission of Kes~79 is highly filamentary and clumpy.
In the east, two striking X-ray filaments are distorted and composed of
with a string of small X-ray clumps in scale of less than $15''$ 
($0.5$ pc at a distance of 7.1 kpc).
The V-shaped structure in the southern part appears to be the 
softest X-ray emitter (due to its relatively red color).
In the SNR interior, two jet-like filaments stretch from the southeast
to the northwest and across the CCO.
Near the northwestern end,  there is a patch of hard X-ray emission,
which is discussed in Section~\ref{S:fragment}.
The two compact sources, the magnetar 3XMM~J185246.6+003317 
and the CCO PSR~J1852+0040, are located in the south and the 
center of the SNR, respectively.

The \Spitzer\ $24~\um$ IR image reveals many thin filaments 
inside the remnant which are well matched with the bright X-ray filaments
(see Figure~\ref{f:rgb}b).
Here we use the unsharp mask IR image\footnote{
We first produce a smoothed (with a Gaussian kernel of 6\farcs{26}), 
or unsharp, image as a positive mask.
The mask is then inverted, scaled, and added to the original image.
The resulted image will increase the sharpness or contrast.}
instead of the original emission image in order to highlight the 
sharp structures.

In figure~\ref{f:rgb}d, we compare the distribution of the
high-$\VLSR$ MCs (110--120~\kms; above the system velocity 
105~\kms; green) with that of the SNR's  X-ray (blue) and radio
(red) emission.
A molecular arc delineates the western outer radio shell
and confines the bright X-ray emission in the SNR interior.
Meanwhile, a northeast-south oriented molecular ridge also 
spatially matches the bright X-ray filaments and eastern inner 
radio emission.

\subsubsection{X-Ray Halo}

A faint X-ray halo surrounding the filamentary structures
appears to stretch to the outer radio boundary, as shown in
the 0.3--7.0 keV X-ray image (Figure~\ref{f:rgb}(c)).
The X-ray image was exposure-corrected, smoothed with a Gaussian 
kernel of $3\farcm{3}$, and
overlaid with contours of the 20 cm radio continuum emission.
Here we display the X-ray image with a histogram equalization scale to
highlight the faint emission.
Unlike the filaments, the faint X-ray halo has no evident IR 
counterpart.
The properties of the X-ray halo and the filaments will be
compared in Section~\ref{S:xspec}.

\begin{figure*}[tbh!]
\centering
\includegraphics[angle=0, width=0.8\textwidth]{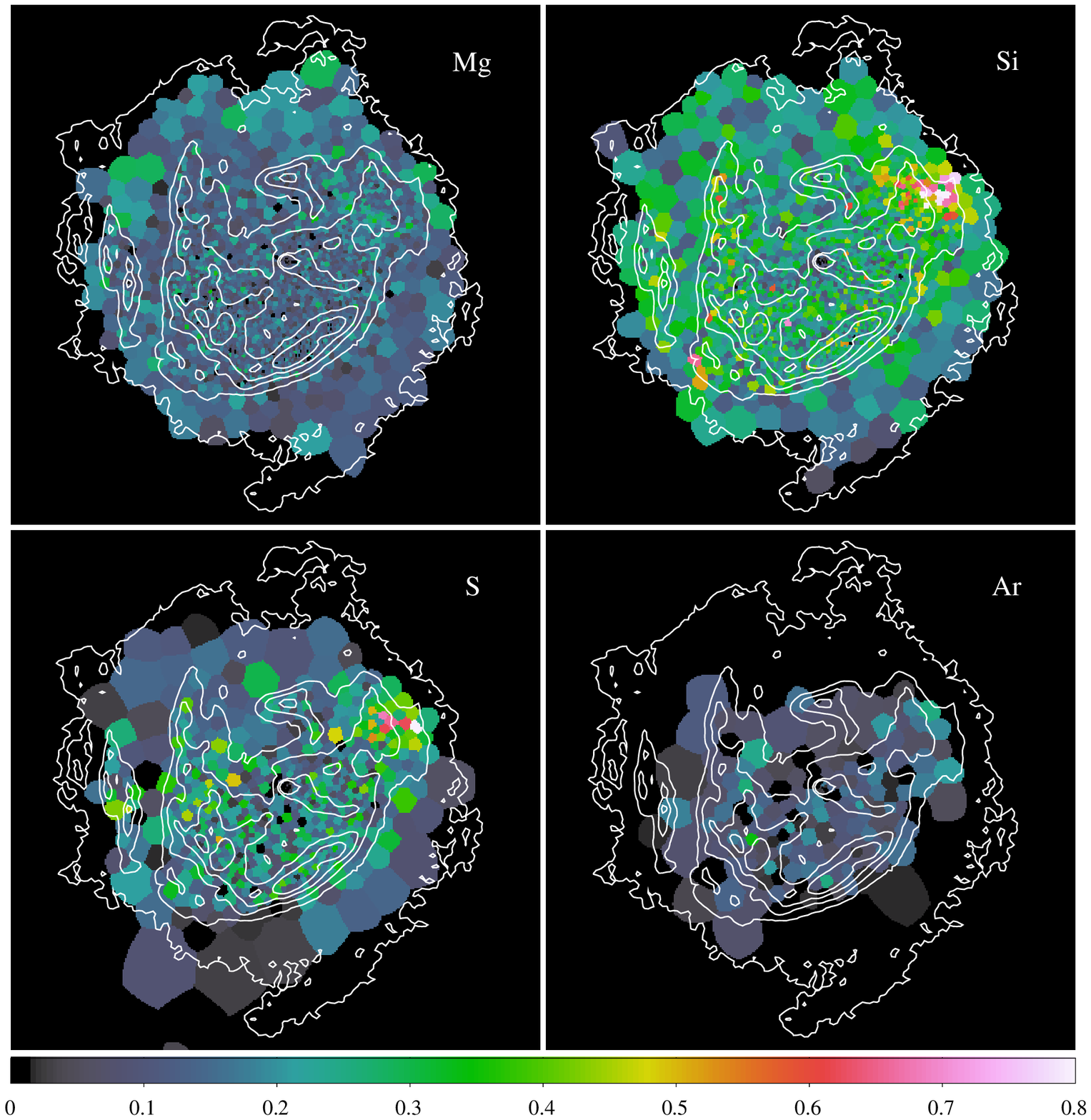}
\caption{EW images of He$\alpha$ emission of Mg, Si, S, and Ar overlaid 
with 0.3--7.0 keV X-ray contours.
The contours are made from the X-ray image as shown in Figure~\ref{f:rgb}c
and the levels are 1.8, 5.6, 11.2, 19.6, and 28 counts.
The band selection is tabulated in Table~\ref{T:ewmap}.
EWs are set to zero for the bins with S/N$<4$ and low continuum fluxes
(4\% of the mean flux for the bright line Mg and Si, and 2\% of the
mean flux for the faint line Si and Ar).
}
\label{f:ewmap}
\end{figure*}

We investigate the distribution of metal lines firstly by using the
EW images of Mg He$\alpha$ at $\sim 1.34$ keV, 
Si He$\alpha$ at $\sim 1.85$ keV, S He$\alpha$ at $\sim 2.45$ keV, 
and Ar He$\alpha$ at $\sim 3.1$ keV (see Figure~\ref{f:ewmap}).
The EW of an emission line is defined as 
$EW=\int I_{\rm l}(E)dE/I_{\rm c}$, where $\int I_{\rm l}(E)dE$ and 
$I_{\rm c}$ are the energy-integrated intensity of the line component 
and the intensity of the underlying continuum at the line center, respectively.
The EW values of metal lines depend on the abundances, and are 
also affected by the temperature and ionization states.

We define the energy bands of the lines, and the left and right 
continuum shoulders as shown in Table~\ref{T:ewmap}, according 
to the global spectra of \snr\ (see Section~\ref{S:result_global} below). 
The continuum emission of each line is estimated by 
interpolating the left and right shoulders' emission.
The background of each image is subtracted before producing the EW map.
The quiescent particle background (QPB) is first subtracted from each 
image since it is apparently spatially variant at $\sim 1.49$~keV
and $\sim 1.74$~keV, which would affect the EWs of the Mg and Si lines.
The QPB images are created from the filter-wheel closed (FWC) data 
with the XMM-ESAS software.
We then estimate the local background level from a region near the 
remnant and subtract it from the QPB-subtracted image of each band.
We adaptively bin the background-subtracted EW images with
signal-to-noise ratio (S/N) $\approx 5$ 
by applying the weighted Voronoi tessellations binning algorithm 
(Diehl \& Statler 2006), which is a generalization of Cappellari 
\& Copin's (2003) Voronoi binning algorithm.
The bins with S/N$<4$ and low continuum fluxes were set to zero.

\subsubsection{Equivalent Width (EW) Images of Metal Species} \label{S:ewmap}

The EWs of the Mg line are nearly uniform across the remnant.
Here the Mg He$\alpha$ line's soft shoulder is strongly affected by 
the variation of the interstellar absorption, which could also bring 
uncertainties to the estimation of the continuum.

Asymmetric distribution is clearly revealed in
both EW images of Si and S.
The largest EW values are found in a patch in the northwest.
Generally, the EWs at the X-ray filaments are larger than those
in the faint halo regions.
A similar trend can also be found in the EW map of Ar.
The asymmetric EW distribution of the metal lines suggests 
that the hot gas properties (abundances, temperature, or 
ionization states) are not uniform in the SNR. 
The halo and filamentary regions may not be in the same physical 
states, as confirmed below according to the spectral analysis.

\subsection{{\rm XMM-Newton} Spectral Analysis} \label{S:xspec}

\subsubsection{Spectral Extraction and Background Subtraction} 

\begin{figure}[t!]
\centering
\includegraphics[angle=0, width=0.5\textwidth]{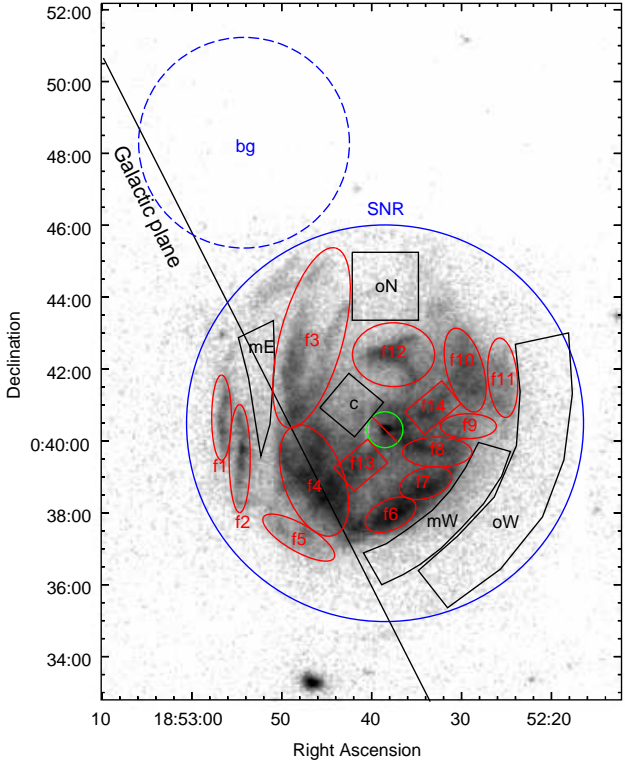}
\caption{EPIC-MOS raw image of \snr\ smoothed with a Gaussian kernel of 3. 
The solid regions colored in blue, red, and black are defined for spectral 
extraction of the whole SNR, the filaments, and diffuse gas, respectively.
The dashed circle labeled ``bg'' is selected as the background region.
The point sources removed in the concerned area are denoted by the green region
with a red slash.
}
\label{f:reg}
\end{figure}

The $\gtrsim$ 375 ks \XMMN\ observations allow us to quantitatively 
investigate the hot gas in small-scale regions in \snr.
We define source (on-SNR) and background (off-SNR) regions for
spectral analysis after removing the central bright X-ray source
PSR J1852+0040 by excluding a circular region with a radius 
of $30''$ centered at ($\RA{18}{52}{38}522$, $\decl{00}{40}{20}89$, 
J2000) (see Figure~\ref{f:reg}).
We select a large region with a radius of $5\farcm{5}$ that covers 
the whole SNR, 14 small regions for 
the bright filamentary structures (``f1--14'') and 5 smaller regions 
(``c'', ``mE'', ``mW'', ``oW'', and ``oN'') for the faint halo gas.
The background region ``bg'' is selected from the nearby sky with
the same Galactic latitude as that of the remnant to minimize 
contamination by the Galactic ridge emission.

\begin{center}
\begin{deluxetable}{p{2.cm}ccc}
\tabletypesize{\footnotesize}
\tablecaption{Energy ranges selected for the EW maps}
\tablewidth{0pt}
\tablehead{
\colhead{Metal Line} & Left Shoulder & Line & Right Shoulder\\
& (keV) & (keV) & (keV) 
}
\startdata
Mg He$\alpha$ & 1.16--1.22 & 1.25--1.42 & 1.45--1.51 \\
Si He$\alpha$ & 1.45--1.51 & 1.75--1.95 & 1.96--2.08 \\
S  He$\alpha$ & 1.96--2.08 & 2.30--2.60 & 2.63--2.78 \\
Ar He$\alpha$ & 2.63--2.78 & 2.98--3.22 & 3.28--3.70
\enddata
\label{T:ewmap}
\end{deluxetable}
\end{center}

We apply a double-subtraction method that takes into account vignetting
effects and the spatially variable instrumental background as follows:
(1) subtract respective instrumental background contribution 
from the raw on- and off-SNR spectra by using the FWC data of MOS1;
(2) construct a background model by fitting the off-SNR spectra
(instrumental background-subtracted), which is subsequently normalized
by area and added to the source model to describe the on-SNR spectra.
FWC data are selected in the epochs close to those of the source
observations in order to reduce variation of instrumental background.
Using the task $addascaspec$ in HEASOFT, we merged the spectra taken within 
two months to increase the statistical quality for the hard X-ray band.
Hence, the spectra of the two 2004 observations, five 2008 observations,
and six 2009 observations are merged respectively.
Finally, the 10 source spectra (5 MOS1 plus 5 MOS2 spectra taken 
from the 2004, 2006, 2007, 2008, and 2009 observations) for each on-SNR
region are then instrumental background-subtracted and binned to reach
an S/N of $> 4$ in the following joint-fit.
We only use the six off-SNR spectra (region ``bg''; 3 MOS1 plus 3 MOS2)
taken in 2004, 2006 and 2008 to construct the background model,
since the region ``bg'' was at the unusable CCD of MOS1 during the
observations in 2007 and 2009.

\begin{figure}[tbh!]
\centering
\includegraphics[angle=270, width=0.5\textwidth]{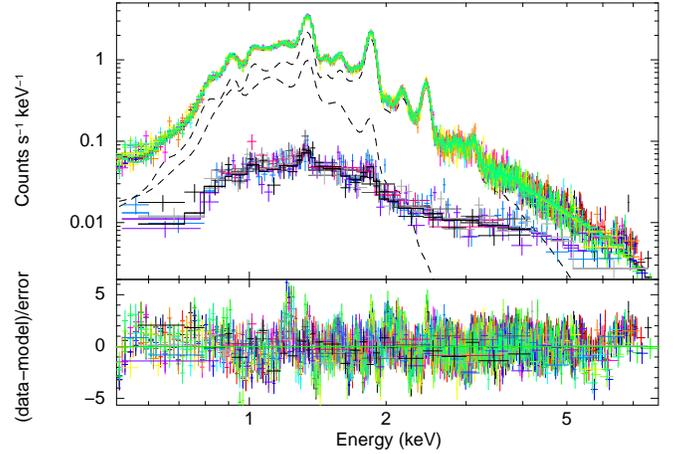}
\vspace{0.2in}
\caption{EPIC-MOS spectra of the whole SNR (upper; 10 groups) and
region ``bg'' (bottom; 6 groups).
The source spectra are fitted with an absorbed \nei\ +\vnei\ 
model (dashed lines) plus the background model (dotted lines; 
phenomenologically describing the ``bg'' spectra; note that the area 
response is different at the the on-SNR and off-SNR regions).
The spectral fitting results are summarized in Table~\ref{T:nei_vnei}.
}
\label{f:snrspec}
\end{figure}

XSPEC (version 12.9) is used for spectral fitting.  
The spectra in the off-SNR region ``bg'' is phenomenologically 
described with an absorbed $nei$+$power$-$law$ plus an unabsorbed 
$bremsstrahlung$ + $gaussian$ model ($\chi_\nu^2/d.o.f=1.33/231$;
see Figure~\ref{f:snrspec}).
The on-SNR sky background is determined by normalizing the model 
according to the region sizes.

\subsubsection{The Global Spectra}  \label{S:result_global}
Figure~\ref{f:snrspec} shows the spectra extracted from source region
``SNR''  (upper) and background region ``bg'' (bottom; normalized by area).
The global spectra of the SNR in 0.5--8 keV reveals distinct He-like 
lines of Mg ($\sim 1.34$ keV), Si ($\sim 1.85$ keV and $\sim 2.19$ keV), 
S ($\sim 2.45$ keV), and Ar ($\sim 3.1$ keV), and several faint H-like lines 
of Ne at 0.9 keV and S at 2.9 keV.

We first applied an absorbed non-equilibrium ionization (NEI) model 
(plus the background model with fixed values) to jointly fit the 10 
source spectra in 0.5-8.0 keV.
For the foreground absorption, we used the $tbabs$ model with
the Anders \& Grevesse (1989) abundances and photoelectric 
cross-section from Balucinska-Church \& McCammon (1992).
The NEI model \vnei\ (NEI version 3.0 in XSPEC) is firstly applied, 
with the abundances of Ne, Mg, Si, S, Ar, and Fe (the abundance of Ni 
is tied to Fe) in the models allowed to vary in the spectral fit.
Here \vnei\ describes an NEI plasma with single and uniform
ionization parameters.
However, the single NEI model fails to describe the spectra in
the hard X-ray band ($\chi_{\nu}^2/d.o.f =2.98/2657$) 
as also pointed out by A14.
The elevated tail of the observed spectra compared to the model above 
3 keV is also 
seen in single thermal model spectral fittings in previous studies 
(e.g. S04, and Giacani \etal\ 2009), implying that the 
global spectrum of \snr\ should contain more than one component.

Since some thermal composite SNRs show recombining plasma,
we tried an absorbed recombining plasma model ($vrnei$ in XSPEC)
to fit the spectra.
However, the spectral fit ($\chi_{\nu}^2/d.o.f=2.96/2656$) gives an initial 
plasma temperature (0.12~keV) much smaller than the current plasma 
temperature (0.75~keV), suggesting that the gas is under-ionized 
rather than over-ionized.

An absorbed \nei+\vnei\ model is subsequently used, which substantially
improves the spectral fit
($\chi_{\nu}^2 /d.o.f =1.92/2654$; see Table~\ref{T:nei_vnei} for 
the spectral fit results and Figure~\ref{f:snrspec} for the spectra).
The abundance of the cool component is set to solar 
since variable abundance does not apparently improve the spectral fit.
The model is still not good enough to reproduce the spectra, especially
in the 1-3~keV band, which suggests that the physical properties of the 
X-ray emitting gas  are not uniform across the remnant.
However, the two-temperature model, to some extent, provides us
with general properties of the hot plasma.

\begin{turnpage}
\tabletypesize{\scriptsize}
\setlength{\tabcolsep}{1.5pt}
\begin{deluxetable*}{lccccccccccccccc}
\tablecaption{Spectral Fitting results with the \nei/\apec+\vnei\ model}
\tablewidth{0pt}
\tablehead{
\colhead{Region}
&$\chi_{\nu}^{2}$/dof & $\NH$ & $\kTc$ & $\tauc$ & $\kTh$ 
& $\tauh$ & Ne & Mg & Si & S & Ar & Fe 
& $n_{\rm c}$ & $n_{\rm h}$
& $t_{i,{\rm h}}$ \\
& & ($10^{22} \cm^{-2}$) & (keV) & ($10^{11} \cm^{-3}$ s)
& (keV) & ($10^{10} \cm^{-3}$ s) & & & & & & 
& ($\cm^{-3}$) & ($\cm^{-3}$) & (kyr)
}
\startdata
SNR
 &        1.92 /2654
 &        1.707$_{      -0.099}^{+       0.004}$
 &        0.195$_{      -0.002}^{+       0.003}$
 &        6.4$\pm 0.4$
 &        0.80$\pm 0.01$
 &        8.1$\pm 0.1$
 &        1.83$_{      -0.05}^{+       0.04}$
 &        1.56$\pm 0.03$
 &        1.40$\pm 0.02$
 &        1.72$\pm 0.03$
 &        1.8$\pm 0.1$
 &        0.93$\pm 0.04$
 & 2.1
 & 0.5
 & 4.2
\\ \hline 
f1
 &        1.01 / 501
 &        1.8$\pm 0.1$
 &        0.21$_{      -0.01}^{+       0.03}$
 &        6.4 ($>2.7$) 
 &        0.7$_{      -0.1}^{+       0.2}$
 &        6.6$_{      -2.3}^{+       4.2}$
 &        2.0$_{      -0.7}^{+       1.3}$
 &        1.8$_{      -0.4}^{+       0.7}$
 &        1.5$\pm 0.3$
 &        1.8$_{      -0.4}^{+       0.7}$
 &        1 (fixed)
 &        1.5$_{      -0.5}^{+       0.9}$
 & 5.7
 & 1.7
 & 1.0
\\
f2
 &        1.06 / 697
 &        1.77$\pm 0.09$
 &        0.19$_{      -0.02}^{+       0.02}$
 &        6.7 ($>2.6$)  
 &        0.73$_{      -0.07}^{+       0.09}$
 &        7.7$_{      -2.0}^{+       2.7}$
 &        2.5$_{      -0.7}^{+       0.9}$
 &        1.6$\pm 0.3$
 &        1.6$_{      -0.2}^{+       0.3}$
 &        2.1$\pm 0.3$
 &        1 (fixed)
 &        1.2$_{      -0.4}^{+       0.5}$
 & 6.0
 & 1.6
 & 1.3
\\
f3
 &        1.19 /1249
 &        1.84$\pm 0.05$
 &        0.20$\pm 0.01$
 &        6.2$_{      -2.2}^{+       3.4}$
 &        0.75$\pm 0.04$
 &        7.6$_{      -1.1}^{+       1.2}$
 &        2.2$\pm 0.03$
 &        1.6$\pm 0.02$
 &        1.6$_{      -0.1}^{+       0.2}$
 &        1.9$\pm 0.2$
 &        2.8$\pm 0.8$
 &        1.1$_{      -0.2}^{+       0.3}$
 & 3.8
 & 1.0
 & 2.0
\\
f4
 &        1.28 /1622
 &        1.63$_{      -0.04}^{+       0.02}$
 &        0.197$_{      -0.004}^{+       0.003}$
 &        5.4$_{      -0.8}^{+       2.7}$
 &        0.91$_{      -0.02}^{+       0.03}$
 &        6.4$_{      -0.6}^{+       0.3}$
 &        2.5$\pm 0.2$
 &        1.8$\pm 0.1$
 &        1.48$_{      -0.08}^{+       0.03}$
 &        1.82$_{      -0.10}^{+       0.07}$
 &        1.5$\pm 0.3$
 &        0.94$_{      -0.13}^{+       0.07}$
 & 8.1
 & 1.8
 & 1.0
\\
f5
 &        0.97 / 614
 &        1.6$\pm 0.1$
 &        0.19$\pm 0.01$
 &        7.5 ($> 2.9$) 
 &        1.4$_{      -0.2}^{+       0.3}$
 &        2.3$_{      -0.4}^{+       0.5}$
 &        2.8$_{      -0.9}^{+       1.1}$
 &        1.9$_{      -0.4}^{+       0.5}$
 &        1.7$_{      -0.3}^{+       0.4}$
 &        1.8$_{      -0.3}^{+       0.4}$
 &        1 (fixed)
 &        1.1$_{      -0.5}^{+       0.6}$
 & 6.2
 & 0.8
 & 0.8
\\
f6
 &        1.18 /1093
 &        1.58$\pm 0.05$
 &        0.200$_{      -0.003}^{+       0.008}$
 &        7.2$_{      -2.0}^{+       7.5}$
 &        0.82$_{      -0.05}^{+       0.06}$
 &        9.6$_{      -1.5}^{+       1.6}$
 &        2.6$\pm 0.5$
 &        1.9$\pm 0.2$
 &        1.7$_{      -0.2}^{+       0.1}$
 &        2.0$\pm 0.2$
 &        1.6$\pm 0.5$
 &        1.0$\pm 0.2$
 & 13.0
 & 3.2
 & 0.8
\\
f7
 &        1.14 /1075
 &        1.68$_{      -0.05}^{+       0.07}$
 &        0.20$\pm 0.01$
 &       13.4 ($>6.5$) 
 &        0.91$_{      -0.04}^{+       0.05}$
 &        7.2$_{      -1.3}^{+       0.9}$
 &        2.3$_{      -0.5}^{+       0.6}$
 &        1.8$\pm 0.2$
 &        1.5$\pm 0.2$
 &        1.8$\pm 0.2$
 &        1.1$\pm 0.5$
 &        0.9$\pm 0.2$
 & 13.2
 & 2.8
 & 0.7
\\
f8
 &        1.20 /1060
 &        1.81$_{      -0.04}^{+       0.03}$
 &        0.21$_{      -0.02}^{+       0.01}$
 & \ldots
 &        0.91$\pm 0.03$
 &        6.8$_{      -0.8}^{+       1.1}$
 &        2.1$_{      -0.3}^{+       0.4}$
 &        2.0$\pm 0.2$
 &        1.6$\pm 0.2$
 &        1.9$\pm 0.2$
 &        1.3$_{      -0.5}^{+       0.6}$
 &        1.1$_{      -0.2}^{+       0.3}$
 & 8.8
 & 2.2
 & 0.9
\\
f9
 &        1.07 / 620
 &        1.9$\pm 0.1$
 &        0.21$\pm 0.03$
 &        7.0 ($>3.5$ ) 
 &        0.9$\pm 0.1$
 &        6.7$_{      -1.5}^{+       3.1}$
 &        2.0$_{      -0.7}^{+       1.1}$
 &        2.2$\pm 0.5$
 &        1.7$_{      -0.3}^{+       0.4}$
 &        1.9$_{      -0.4}^{+       0.5}$
 &        2.6$_{      -1.2}^{+       1.4}$
 &        1.5$_{      -0.5}^{+       0.7}$
 & 7.6
 & 1.8
 & 1.0
\\
f10
 &        1.17 /1064
 &        1.90$\pm 0.03$
 &        0.22$\pm 0.01$
 & \ldots
 &        1.18$_{      -0.06}^{+       0.07}$
 &        2.4$\pm 0.2$
 &        2.8$_{      -0.5}^{+       0.6}$
 &        2.8$_{      -0.3}^{+       0.4}$
 &        3.1$\pm 0.3$
 &        3.4$\pm 0.4$
 &        3.6$_{      -1.1}^{+       1.2}$
 &        1.0$\pm 0.3$
 & 6.6
 & 1.2
 & 0.5
\\
f11
 &        1.03 / 629
 &        2.0$\pm 0.1$
 &        0.19$\pm 0.01$
 &        5.9$_{      -2.7}^{+       9.8}$
 &        1.6$_{      -0.3}^{+       0.6}$
 &        3.2$_{      -0.7}^{+       0.8}$
 &        4.4$_{      -2.2}^{+       3.5}$
 &        3.6$_{      -1.0}^{+       1.7}$
 &        5.2$_{      -1.3}^{+       2.2}$
 &        5.2$_{      -1.1}^{+       1.8}$
 &        4.2$_{      -1.8}^{+       2.4}$
 &        0.6$_{      -0.6}^{+       1.6}$
 & 7.9
 & 1.0
 & 0.9
\\
f12
 &        1.11 /1066
 &        1.82$_{      -0.06}^{+       0.05}$
 &        0.20$_{      -0.01}^{+       0.02}$
 &        7.5$_{      -3.4}^{+      20.3}$
 &        0.74$_{      -0.04}^{+       0.05}$
 &        9.9$_{      -1.8}^{+       2.0}$
 &        1.8$\pm 0.4$
 &        1.7$\pm 0.2$
 &        1.4$\pm 0.2$
 &        1.8$\pm 0.2$
 &        1.2$_{      -0.6}^{+       0.7}$
 &        1.1$_{      -0.2}^{+       0.3}$
 & 6.0
 & 1.7
 & 1.6
\\
f13
 &        1.07 / 998
 &        1.66$\pm 0.04$
 &        0.18$\pm 0.01$
 & \ldots
 &        0.89$_{      -0.06}^{+       0.05}$
 &       10.0$_{      -2.0}^{+       1.9}$
 &        2.1$\pm 0.3$
 &        1.9$\pm 0.2$
 &        1.5$_{      -0.1}^{+       0.2}$
 &        1.6$\pm 0.2$
 &        1.7$\pm 0.5$
 &        1.0$_{      -0.2}^{+       0.3}$
 & 12.4
 & 2.5
 & 1.0
\\
f14
 &        1.10 / 924
 &        1.83$_{      -0.04}^{+       0.03}$
 &        0.22$\pm 0.01$
 & \ldots
 &        0.88$_{      -0.10}^{+       0.07}$
 &        7.2$_{      -1.1}^{+       2.2}$
 &        2.3$\pm 0.5$
 &        1.9$_{      -0.2}^{+       0.3}$
 &        1.6$_{      -0.2}^{+       0.1}$
 &        1.8$_{      -0.2}^{+       0.3}$
 &        1.2$_{      -0.6}^{+       0.7}$
 &        0.7$\pm 0.3$
 & 9.1
 & 2.2
 & 0.9
\\ \hline 
c
 &        0.98 / 723
 &        1.9$\pm 0.1$
 &        0.19$\pm 0.01$
 &        4.8$_{      -2.4}^{+       5.7}$
 &        0.9$\pm 0.1$
 &        7.8$_{      -1.6}^{+       2.3}$
 &        3.0$_{      -0.8}^{+       1.5}$
 &        1.5$\pm 0.3$
 &        1.3$_{      -0.2}^{+       0.3}$
 &        1.6$_{      -0.2}^{+       0.3}$
 &        0.9$_{      -0.7}^{+       0.8}$
 &        0.5$_{      -0.5}^{+       0.4}$
 & 2.4
 & 0.5
 & 4.1
\\
mE
 &        1.15 / 351
 &        1.62$_{      -0.08}^{+       0.11}$
 &        0.17$\pm 0.05$
 & \ldots
 &        0.7$_{      -0.1}^{+       0.3}$
 &        7.2$_{      -4.5}^{+      10.4}$
 &        0.9$_{      -0.6}^{+       0.9}$
 &        1.2$_{      -0.3}^{+       0.5}$
 &        1.1$_{      -0.2}^{+       0.4}$
 &        1.3$_{      -0.4}^{+       0.5}$
 &        1 (fixed)
 &        1.1$_{      -0.4}^{+       0.7}$
 & 1.4
 & 0.3
 & 6.3
\\
mW
 &        1.15 / 836
 &        1.66$_{      -0.08}^{+       0.06}$
 &        0.20$_{      -0.01}^{+       0.02}$
 &        7.8 ($>4.4$) 
 &        1.0$_{      -0.1}^{+       0.2}$
 &        5.4$_{      -1.6}^{+       1.9}$
 &        2.3$_{      -0.6}^{+       1.1}$
 &        1.8$_{      -0.3}^{+       0.4}$
 &        1.5$_{      -0.2}^{+       0.3}$
 &        1.7$_{      -0.3}^{+       0.4}$
 &        1.6$_{      -1.1}^{+       1.2}$
 &        1.1$\pm 0.4$
 & 2.1
 & 0.4
 & 3.5
\\
oW
 &        1.07 / 804
 &        1.72$\pm 0.08$
 &        0.18$_{      -0.01}^{+       0.02}$
 &        2.1$_{      -1.2}^{+       3.8}$
 &        0.68$_{      -0.06}^{+       0.09}$
 &       12.4$_{      -5.1}^{+      10.8}$
 &        1.3$_{      -0.4}^{+       0.6}$
 &        1.4$_{      -0.2}^{+       0.4}$
 &        1.2$_{      -0.2}^{+       0.4}$
 &        1.5$_{      -0.3}^{+       0.5}$
 &        1 (fixed)
 &        1.7$_{      -0.5}^{+       0.7}$
 & 1.2
 & 0.3
 & 10.9
\\
oN
 &        1.15 / 406
 &        1.7$\pm 0.1$
 &        0.19$_{      -0.03}^{+       0.04}$
 &        4.5 ($>1.1$) 
 &        0.7$_{      -0.1}^{+       0.2}$
 &        7.6$_{      -3.3}^{+       8.4}$
 &        1.9$_{      -0.8}^{+       1.6}$
 &        1.8$_{      -0.5}^{+       1.0}$
 &        1.5$_{      -0.4}^{+       0.8}$
 &        1.7$_{      -0.6}^{+       1.0}$
 &        1 (fixed)
 &        1.3$_{      -0.7}^{+       1.1}$
 & 1.8
 & 0.5
 & 4.0
\enddata
\label{T:nei_vnei}
\tablecomments{
The errors are estimated at the 90\% confidence level.
}
\end{deluxetable*}
\end{turnpage}

According to the \nei+\vnei\ model, the X-rays from the remnant 
suffers an interstellar absorption $\NH=1.71 \E{22} \cm^{-2}$ and
the X-ray-emitting plasma can be roughly described as under-ionized
two-temperature gas.
The cool phase of the gas has a temperature $\kTc= 0.19$~keV, 
an ionization timescale 
$\tauc=n_{e,{\rm c}} t_{i,{\rm c}}=6.4\pm0.4\E{11}~\cm^{-3}\s$ , 
and solar abundances (so fixed to 1), 
while the hot phase has a temperature $\kTh= 0.80$~keV, an ionization 
timescale $\tauh=n_{e,{\rm h}} t_{i,{\rm h}}=8.1\pm0.1\E{10}~\cm^{-3}\s$
and enriched metal abundances ([Ne]/[Ne]$_\odot=1.83\pm0.05$,
[Mg]/[Mg]$_\odot=1.56\pm 0.03$, [S]/[S]$_\odot=1.40\pm 0.02$,
and [Ar]/[Ar]$_\odot=1.8\pm0.1$).

The column density (inferred from the $tbabs$ model) is larger than that 
obtained by S04
($1.54\pm0.5\E{22}~\cm^{-2}$, $phabs$ model) and Giacani \etal\ (2009; 
$1.52\pm0.02\E{22}~\cm^{-2}$, $phabs$ model), but smaller than that
reported by A14 ($2.43\pm0.05\E{22}~\cm^{-2}$; $tbabs$ model).
Some other spectral fit results, such as $\kTc$ and abundances,
are also different from those obtained by A14.
We believe that the background selection is the main reason for the 
discrepancy between our work and A14.
In particular, some of the background regions selected by A14 are well 
inside the SNR where halo X-ray emission is present 
(see Figure~\ref{f:rgb}c in this paper in comparison to Figure~3 in A14).

We assume that the two-temperature gas fills the whole volume 
($f_{\rm c}+ f_{\rm h}=1$) and is in pressure balance 
($n_{\rm c}T_{\rm c} =n_{\rm h}T_{\rm h}$), where
$f$ and $n$ are the filling factor and hydrogen density ($n=1.2\nel$), 
respectively.  The subscripts ``c'' and ``h'' denote the parameters for the 
cool and hot phases, respectively.
The \nei+\vnei\ fit results (see Table~\ref{T:nei_vnei}) give an 
emission measure $f_{\rm c} n_{e,{\rm c}} n_{\rm c} V \du^{-2}
=5.9^{+0.3}_{-0.2}\E{59}~\cm^{-3}$ for the cool phase gas, 
and $f_{\rm h} n_{e,{\rm h}} n_{\rm h} V \du^{-2}
=2.2\pm0.1\E{58}~\cm^{-3}$ for the hot phase,
where $V$ is the volume of the sphere and $\du$ is the 
distance scaled to 7.1~kpc.
The cool component is found to fill 62\% of the volume, with a 
larger density 
$n_{\rm c}=2.1^{+0.5}_{-0.3}(f_{\rm c}/0.62)^{-1/2}\du^{-1/2}~\cm^{-3}$,
while the hot component has a density $n_{\rm h}= 
0.51^{+0.02}_{-0.01} (f_{\rm h} /0.38)^{-1/2}\du^{-1/2}~\cm^{-3}$.
Then we can obtain the ionization age for the two components
$t_c=\tauc/n_{\rm e,c}\sim 8.1$~kyr and $t_h=\tauh/n_{\rm e,h}
\sim 4.2$~kyr.
The larger ionization age of the cool phase is possibly due to
a variable $n_{\rm e,c}$ (e.g., decreasing with time).

\subsubsection{Spatially resolved spectroscopy}\label{S:spatial}

The X-ray image of \snr\ (see Figure~\ref{f:rgb}a) has revealed a
non-uniform nature of the hot gas.
The EW maps of metal lines (Figure~\ref{f:ewmap}) also 
indicate a nonuniform distribution of the ejecta.
The gas properties are thus expected to be spatially variant, and 
the global spectra may not be well described with a simple one-component 
or two-component model.
We thus compare the spectral analysis results of the 19 defined small 
regions to investigate the variation of gas properties across the SNR.

A two-component thermal model is needed to describe the spectra well 
for the 19 small-scale regions.
Adding an extra model to a single thermal model is also feasible
according to the {\sl F}-test null hypothesis probabilities.
We first use \nei (soft)+ \vnei (hard) to fit the spectra, and
then replace \nei\ with \apec\ (APEC thermal model describing the plasma
in collisional ionization equilibrium; ATOMDB version 3.0.2) for the cool 
phase if its ionization timescale reaches higher than $\sim10^{13}~\cm^{-3}\s$.
Here we allow the Ne, Mg, Si, S, Ar, and Fe abundances of
the hot component to vary as long as they can be well constrained.
The abundance of the cool component is set to solar 
since variable abundance does not apparently improve the spectral fit.
The best-fit results for the \nei/\apec+\vnei\ model are tabulated in
Table~\ref{T:nei_vnei}.\footnote{
We also tried the \pshock/\apec+\vpshock\ model for a comparison, 
where \pshock\ is a plane-parallel shock plasma model with 
a range of ionization parameters and solar abundances, 
while \vpshock\ has variable abundances.
The two models result in almost identical temperature and abundance 
values, with larger $\NH$ and $\tau_h$ obtained with the latter model.
The larger $\tau_h$ value is interpretable given that in \vpshock\ 
it stands for the upper limit of the ionization scale while in \vnei\
a single ionization scale is adopted.
In this paper, we use the fit results of the \nei/\apec+\vnei\ model 
because it is based on a simpler uniform assumption and generally 
provides a good characterization of the gas's general properties.}

The interstellar column density $\NH$ of the X-ray emission varies from 
$\sim 1.5\E{22}
\cm^{-2}$ in the south to $\sim 2.0\E{22} \cm^{-2}$ in the northwest, 
consistent with the fact that the southern portion emits softer X-rays
(as shown in Figure~\ref{f:rgb}a).
Generally, both the bright filamentary (except regions ``f5'', ``f10'' 
and ``f11'') and faint halo gas hold spectral properties similar
to the global SNR gas, including the temperature of the cool phase 
($\kTc\sim 0.2$~keV), and the temperature ($\kTh\sim 0.7$--1.0~keV), 
ionization timescale ($\tauh=6$--$12\E{10}~\cm^{-3} \s$), and
metal abundances of the hot phase.
Nevertheless, there is a trend that the gas near the SNR edge has
lower $\kTh$ ($\sim 0.7$~keV) than in the SNR center ($\sim 0.9$~keV).

To estimate the density of the plasma, we assume several three-dimensional 
shapes for the two-dimensional regions:
(1) oblate spheroids for the elliptical regions ``f1--12'';
(2) oblate cylinders for the rectangular regions ``f13--14'';
(3) sections of spherical layers for the shell-like halo regions
``mW'' and ``oW'';
4) for the halo regions ``c'', ``mE'' and ``oN'', the volumes 
are expressed as $V=S\times l$; where $S$ is the area of the
region and $l$ is the line of sight length along the SNR sphere (depending on
the distance of the region to the SNR center).
We assume that $f_{\rm c}+ f_{\rm h}=1$ and $n_{\rm c}T_{\rm c} 
=n_{\rm h}T_{\rm h}$, similar to the calculation for the global gas.
The calculated densities of the two-phase gas are summarized in 
Table~\ref{T:nei_vnei}.

\section{Discussion} \label{S:discussion}

\subsection{Multi-wavelength Manifestation of the SNR-MC Interaction}

\subsubsection{Molecular environment} \label{S:mol_env}

Using the multi-transition CO observations, we have provided kinematic 
and new morphological evidence to support the scenario that \snr\ is 
interacting with MCs at a systemic LSR velocity of $\sim 105~\km\ps$. 

The column density and mass of the surrounding molecular gas can be 
estimated by using the \thCO~\Jotz\ lines spanning the velocity range
of 95--115~\kms. 
Assuming that the rotational
levels of the \thCO\ molecules are in local thermodynamic equilibrium, 
we obtain the column density of \thCO\ molecules as follows (Kawamura \etal\ 1998):

\begin{equation}
N(^{13}{\rm CO})=2.42\E{14}~\frac{\tau(^{13}{\rm CO}) \Delta v (\km\ps) 
T_{\rm ex} (\K)} {1-\exp(-5.29 \K/T_{\rm ex})}~\cm^{-2},
\end{equation}

where $T_{\rm ex}$ and $\Delta v$ are the excitation temperature 
and line width of \thCO~\Jotz, respectively.
When the \thCO\ lines are optically thin (optical depth 
$\tau(^{13}{\rm CO}) \ll$ 1), the main beam brightness temperature 
of \thCO\ is $T_{\rm mb}\approx T_{\rm ex} \tau(^{13}{\rm CO})$.
Considering that $T_{\rm mb}\Delta v\approx \int T_{\rm mb}(v) dv$
and taking the $\NHH$ to $N(^{13}{\rm CO})$ ratio as $5\E{5}$
(Dickman 1978), we obtain the column density of H$_2$:

\begin{equation}
\NHH=1.21\E{20}\frac{\int T_{\rm mb}(v)dv
(\K)} {1-\exp(-5.29/T_{\rm ex})}~\cm^{-2}.
\end{equation}

\begin{figure}[t!]
\centering
\includegraphics[angle=0, width=0.45\textwidth]{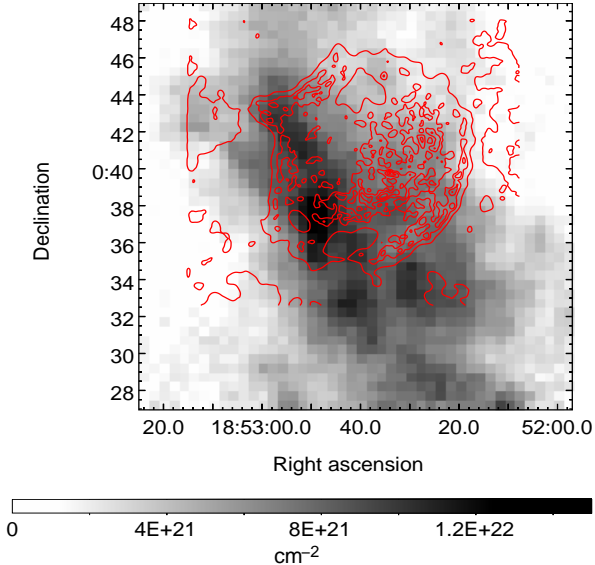}
\caption{
Distribution of the H$_2$ column density (estimated from \thCO~\Jotz\ in
the velocity range 95--115~\kms) in the vicinity of \snr,
VLA 20~cm radio contours are overlaid.
}
\label{f:nh2}
\end{figure}

A typical MCs temperature of 10~K is adopted here for $T_{\rm ex}$,
which is also close to the peak main beam temperature of \twCO~\Jotz\ lines
in the vicinity of \snr\ (see Figure~\ref{f:linegrid}).
The \thCO\ emission integrated in the velocity range 95--115~\kms\ 
(associated with the SNR)  is then used to derive the $\NHH$ and
the distribution of $\NHH$\ near the remnant is revealed in Figure~\ref{f:nh2}.

The mass of the environmental molecular gas in the $22'\times 22'$ 
(or $45.4\du$~pc $\times 45.4\du$~pc) region (field of view in 
Figure~\ref{f:nh2}), is obtained to be 
$M_{\rm mol}=\mu m_{\rm H} \sum[ \NHH  L^2]\sim5.7\du^2\E{4}~\Msun$,
where the mean molecular weight $\mu=2.8$, 
$m_{\rm H}$ is the mass of the hydrogen atom, and $L$ is the size
of the gas ($45.4\du$~pc).
The size and mass values of the cloud is typical for giant MCs (GMCs;
Dame \etal\ 2001).
The mean molecular density of the GMC can be estimated as 
$\nHH\sim{\overline\NHH}/L\sim 30 \du^{-1}~\cm^{-3}$, where
${\overline \NHH}$ is the mean column density.

\subsubsection{Origin of the Mid-IR and X-Ray Filaments} \label{S:Xfilaments}

\begin{figure}[tbh!]
\centering
\includegraphics[angle=0, width=0.45\textwidth]{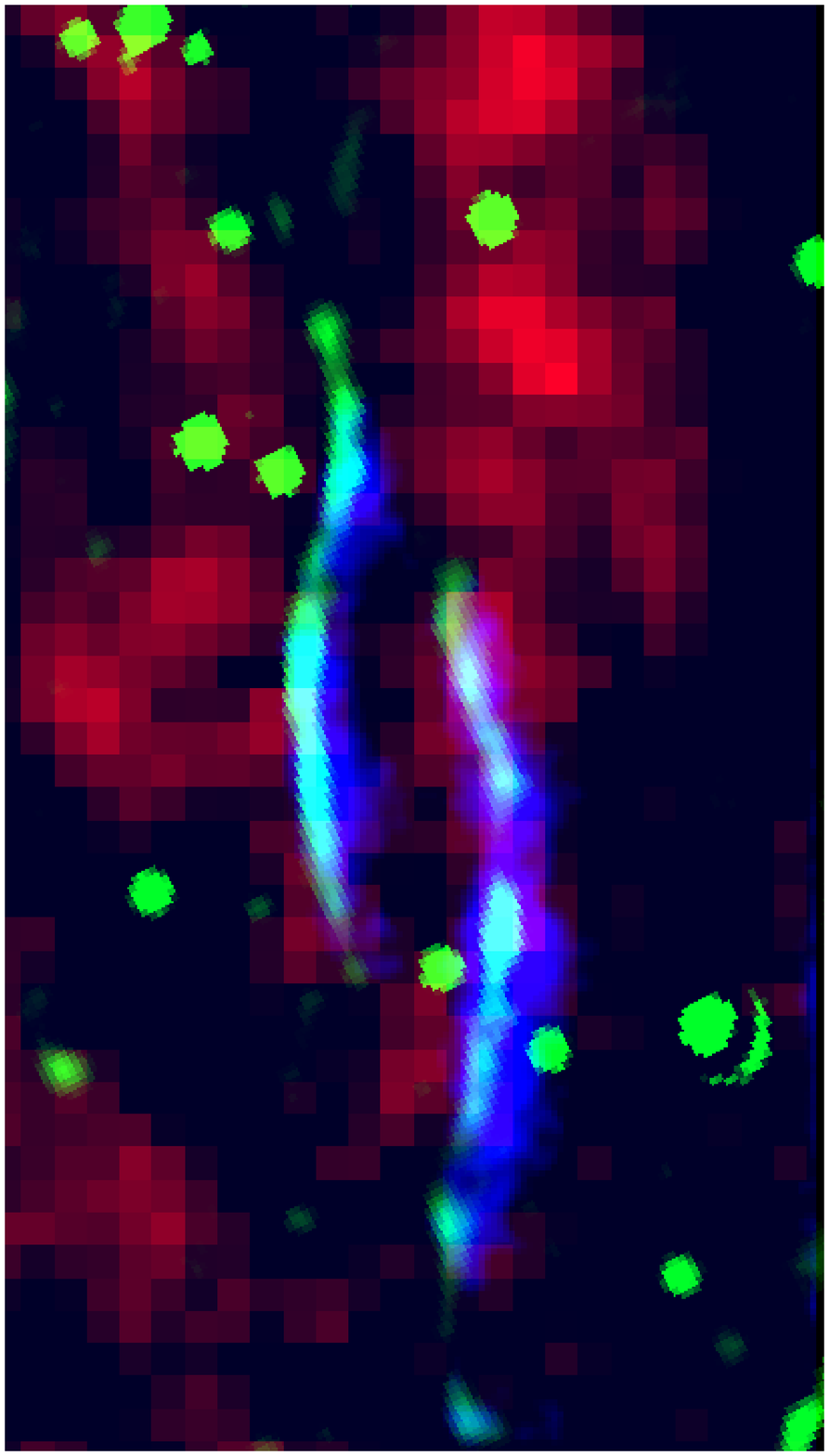}
\caption{Tri-color image of the twin filaments in the east of \snr.
Red: JCMT \twCO~\Jttt\ image at 105~\kms; green: unsharp mask 
\Spitzer\ 24~$\um$ image; blue: \XMMN\ MOS 0.3--7.0 keV.  }
\label{f:twinfilaments}
\end{figure}

A good spatial correlation is present between the mid-IR filaments and
the X-ray filaments (see Figure~\ref{f:rgb}).
Previous IR studies toward middle-age SNRs summarized three main 
origins of the emission at the \Spitzer\ 24~$\um$ band:
(1) continuum emission of dust grains (e.g., Andersen \etal\ 2011); 
(2) atomic lines such as [\ion{Fe}{2}] at 25.988~$\um$
and [\ion{O}{4}] at 25.890~$\um$ (e.g., Reach \& Rho 2000), and 
(3) molecular lines such as the vibrational rotational line of H$_2$ 
at 28.2~$\um$ (e.g., Neufeld \etal\ 2007).
The IR emission at $\gtrsim 15$~$\um$ is usually dominated by the continuum
of dust grains with typical temperature $\sim 40$--100~K (Seok \etal\
2013), while at the shorter wavelengths, the IR emission of SNRs can be 
dominated by ionic/molecular lines (e.g., IC~443, Cesarsky \etal\ 1999; 3C~391,
Reach \etal\ 2002).
The warm dust could be heated by either collisions or radiation.
In the collisions case, the swept-up dust grains are collisionally heated 
by the electrons in the hot plasma (Dwek 1987), which results in a good spatial
correlation between the IR and X-ray emission.
The dust grains could also be heated by the UV photons from the postshock gas 
in the radiative shock (Hollenbach \& McKee 1979), which may be efficient
for the SNRs interacting with MCs.
Considering the MC--IR-X--ray correlation revealed in \snr\ (Figure~\ref{f:rgb}), 
the two cases may both work here.
The SNR has been considered to be in the Sedov phase of evolution (S04; 
see also discussion in Section~\ref{S:global}) and no optical emission 
has been detected to support the radiative shock scenario (could possibly be
due to heavy absorption). 
However, small-scale radiative shock in MCs cannot be excluded.
Since the IR emission is filamentary and a strong IR-X-ray correlation 
is widely seen in the whole remnant
(see Figure~\ref{f:rgb}b),
it is more likely that the $24~\um$ emission comes from dust 
heated by the hot plasma.
Further IR spectroscopic observations are required to provide a firm
conclusion.

The most prominent IR filaments emerge in the eastern section of the SNR, where there is good 
correlation with the X-ray structures, including the enhancements
at several clumps.
In Figure~\ref{f:twinfilaments}, we show the two filaments in the eastern part
of the SNR revealed in \twCO~\Jttt, 24~$\um$, and the 0.3--7.0 keV X-ray bands.
The three bands correspond to three hierachical gas phases: cold dense
gas, warm gas/dust, and hot tenuous plasma.
There is a trend wherein the colder gas is located east of the hotter gas.
In particular, the thin IR filaments with a width of $6''$ (0.2~pc at 
a distance of 7.1~kpc) are just east of the X-ray filaments ($\sim15''$).
The trend is consistent with a scenario where the shock is interacting 
with multi-phase ISM in the east, which then generates 
different types of shocks.
The shock is damped while it propagates into denser medium.
The fast shocks heat the moderate-density cloudlet ($n_{\rm c}\sim 
6~\cm^{-3}$) to a temperature of $\kTc\sim0.2$~keV and the low-density
inter-cloud medium ($n_{\rm h}\sim 1$--$2~\cm^{-3}$) to a temperature of 
$\kTh\sim 0.7$~keV (see Table~\ref{T:nei_vnei}).
The nondissociative shock transmitted into the dense molecular clumps 
($> 10^4 \cm^{-3}$ for the clumps emitting \twCO~\Jttt\ emission)
has a low velocity ($\lesssim 50 \km\ps$, Draine \& Mckee
1993) and generates the broad/asymmetric \twCO\ lines. 
Therefore, the multiple shocks in clumpy ISM explains the bright
filamentary radiation in multi-wavelengths.

The interior X-ray filaments probably arise from 
SNR interaction with relatively dense gas
($\sim 4$--$13~\cm^{-3}$ for the cold component), as compared to
the low density of the halo gas 
($\sim 1$--$2~\cm^{-3}$ for the cold component; see 
Table~\ref{T:nei_vnei}).

\subsubsection{Projection Effects} \label{S:projection}

\snr\ was classified as a mixed-morphology or thermal composite SNR 
(Rho \& Petre 1998), as its brightest X-ray emission arises from 
the interior and the radio emission is shell-like, 
Different from most other thermal composite-type SNRs, which have a smooth, 
centrally filled X-ray morphology, \snr\ has a highly structured 
X-ray morphology, and the interior X-ray and radio emission 
are correlated (see Figure~\ref{f:rgb}).
As S04 note, the inner X-ray shell and outer X-ray halo have 
clear-cut edges and may both come from shock surfaces.
In Section~\ref{S:Xfilaments}, we have provided evidence
that the X-ray and IR filaments, and the radio shells
in the SNR interior result from the interaction of the shock
with a dense medium.
Hence, projection effect is the most plausible origin of the 
multiple-shell structure mixed-morphology nature.

\snr\ may have a double-hemisphere morphology viewed essentially
along the symmetric axis.
SNRs breaking out from an MC into a low-density region can produce such a
double-hemisphere morphology as predicted by hydrodynamic 
simulations (e.g., Tenorio-Tagle \etal\ 1985; Cho \etal\ 2015).
There are a number of SNRs showing such double-hemisphere morphology
due to shock breaking out from a dense medium, such as
IC~443 (e.g., Troja \etal\ 2006), W28 (e.g., Dubner \etal\ 2000), 3C391
(e.g., Reynolds \& Moffett 1993, Chen \etal\ 2004), 
VRO 42.05.01 (e.g., Pineault \etal\ 1987),
G349.7$+0.2$ (e.g., Lazendic \etal\ 2005), etc.
All of these SNRs are associated with MCs (see Jiang \etal\ 2010 and
references therein).
Here we suggest that \snr\ is among the group of double-hemisphere SNRs 
interacting with MCs.
The bright filaments correspond to the smaller hemisphere 
(radius $R_1\sim 4'$) evolving into the denser ISM, while the shock
breaking out in a tenuous medium forms a larger hemisphere 
(radius $R_2\sim 6'$) filled with hot halo gas.
The larger hemisphere or the blow-out part is the blueshifted one 
essentially heading toward us.
This picture can be confirmed based on the distribution of the
foreground absorption, $\NH$, and the molecular line analysis.
As summarized in Table~\ref{T:nei_vnei}, 
$\NH$ is lowest in the southeast (``f5'' and ``mE'') and southwest 
(``f6'' and ``mW'') of the remnant.
Since the column density of MC at these regions can reach as high as 
$\sim 10^{22}~\cm^{-2}$ (see Figure~\ref{f:nh2}), heavy absorption 
would have been present if the MCs were in the foreground. 
Hence, we suggest a picture where most of the molecular gas
(except that in the east) is behind the larger hemisphere (halo gas) 
and the X-ray filaments correspond to the SNR--dense-gas interacting regions.
This scenario is also in agreement with the \twCO\ shell being
red-shifted due to shock perturbation ($\VLSR=110$-120~\kms; see the
detailed description in Section~\ref{S:mol}).

\subsection{Global evolution parameters} \label{S:global}

Most small-scale regions have temperatures $\kTc$ and $\kTh$
similar to those of the global gas (see Table~\ref{T:nei_vnei}), 
while the filling factors $f_{\rm c}$ could be varying across 
the remnant.
Hence, the overall spectral results can represent the average 
properties of the SNR's X-ray-emitting plasma. 
The masses for the cool and hot phases ($M=1.4 \nH \mH V$) are 
$277^{+7}_{-4} (f_{\rm c}/ 0.62)^{1/2}\du^{5/2} ~\Msun$ and $42\pm1 
(f_{\rm h}/0.38)^{1/2} \du^{5/2}~\Msun$, respectively.
Both masses are too high to be produced by the SN ejecta, and are most 
likely due to the shock-heated ISM.
The enriched S and Ar in the hot component indicates that part of 
the mass in the hot phase must be also contributed by the ejecta.
Hence, the hot phase with lower density ($0.5~\cm^{-3}$) and elevated 
metal abundances is related to the emission in the inter-cloud medium, 
while the cool phase with higher density ($2~\cm^{-3}$) and solar 
abundances is suggested to come from the shocked denser cloud.

We use the parameters of the hot phase gas to investigate the
SNR evolution in the inter-cloud medium. 
The bright X-ray emission and low ionization age ($t_i\sim 4.2$~kyr)
for the hot component suggest that \snr\ has not yet entered radiative
phase (at least globally). 
Hence, we adopt the Sedov evolutionary phase (Sedov 1959) for 
\snr\ as done in previous studies of the remnant (e.g. S04 and A14).
Adopting $\kTh=0.80\pm 0.01$~keV as the emission measure
weighted temperature for the whole remnant, the post-shock 
temperature can be estimated as $kT_{\rm s}=\kTh/1.27$
(for ion--electron equipartition; Borkowski \etal\ 2001),
which is $0.63\pm 0.01$~keV.
The shock velocity can be derived as $v_{\rm s}=[16 kT_{\rm s}/(3
\mu m_{\rm H})]^{1/2}=727\pm 5$~\kms, where
$\mu=0.61$ the mean atomic weight for fully ionized plasma.
Taking the curvature radius of the western shell $R_{\rm 2}=12.4\du$~pc
($6'$), we estimate the explosion energy as
$E=(25/4\xi)(1.4n_0 \mH)R_2^3 v_{\rm s}^2
\sim 2.7\E{50} \du^{5/2}$~erg, where $\xi=2.026$ and
the preshock diffuse gas density $n_0=n_{\rm h}/4\sim 0.13~\cm^{-3}$.
The Sedov age of the remnant is estimated as 
$t=2R_s /(5v_{\rm s})\approx6.7$~kyr. 
The real age might deviate from 6.7 kyr to some extent 
because of the non-spherical evolution of \snr. 
Varying $R_s$ between $R_1$ ($\sim 4'$) and $R_2$
($\sim6'$) gives an age range between 4.4--6.7~kyr. 
The lower limit 4.4~kyr is close to the ionization age of the 
hot component (4.2~kyr).

\subsection{High-velocity Ejecta Fragment(s)} \label{S:fragment}

The asymmetric metal distribution in \snr\ is supported by both 
the EW maps of Si and S (see Figure~\ref{f:ewmap}) and the 
spatially resolved spectral results (see Table~\ref{T:nei_vnei}).
Notably, a bright patch is revealed in the Si and S EW maps.
The northwestern regions ``f11'', ``f10'' have a distinctly
higher $\kTh$ and abundances compared to other filamentary regions.
The regions overlap the bright patch in the EW maps of Si and S.

The metal-rich patch mainly corresponds to the spectral
extraction region ``f11'' at $\RA{18}{52}{25}4$, $\decl{00}{41}{45}5$
 (J2000), where the hot component gas has a high $\kTh$ 
($=1.6^{+0.6}_{-0.3}$~keV) and the largest abundances of 
Ne ($4.4^{+3.5}_{-2.2}$), Mg ($3.6^{+1.7}_{-1.0}$), Si ($5.2^{+2.2}_{-1.3}$), 
S ($5.2^{+1.8}_{-1.1}$) and Ar ($4.2^{+2.4}_{-1.8}$). 
The patch is a protrusion of the filaments in the northwest
(see Figure~\ref{f:rgb}a) and may be an ejecta fragment or a 
conglomeration of ejecta fragments at a high velocity.
Using the hot component temperature ($\kTh=1.6^{+0.6}_{-0.3}$~keV), 
the velocity of the fragment is estimated to be
$v_{\rm ej}=[16 \kTh/(3 \mu m_{\rm H})]^{1/2}=1.0^{+0.2}_{-0.1}\E{3}$\kms,
which is 41\% faster than the mean shock velocity $\sim 730$~\kms\ derived
in Section~\ref{S:global}.
In the opposite direction, the plasma in region ``f5'' has a high
temperature ($\kTh=1.4^{+0.3}_{-0.2}$~keV; about twice the average value)
and a lower ionization timescale ($\tauh=2.3^{+0.5}_{-0.4}\E{10} 
\cm^{-3} \s$) in the hot component,
while the metal abundances are not significantly elevated.
The gas in region ``f5'' thus could be fast ejecta clump well mixed 
with the ambient medium.

The non-uniform distribution of metal species in \snr\ 
and the presence of high-velocity ejecta fragement(s) reflect
the intrinsic asymmetries of the SN explosion.

\subsection{Constraints on the progenitor}

\begin{figure}[tbh!]
\centering
\includegraphics[angle=0, width=0.5\textwidth]{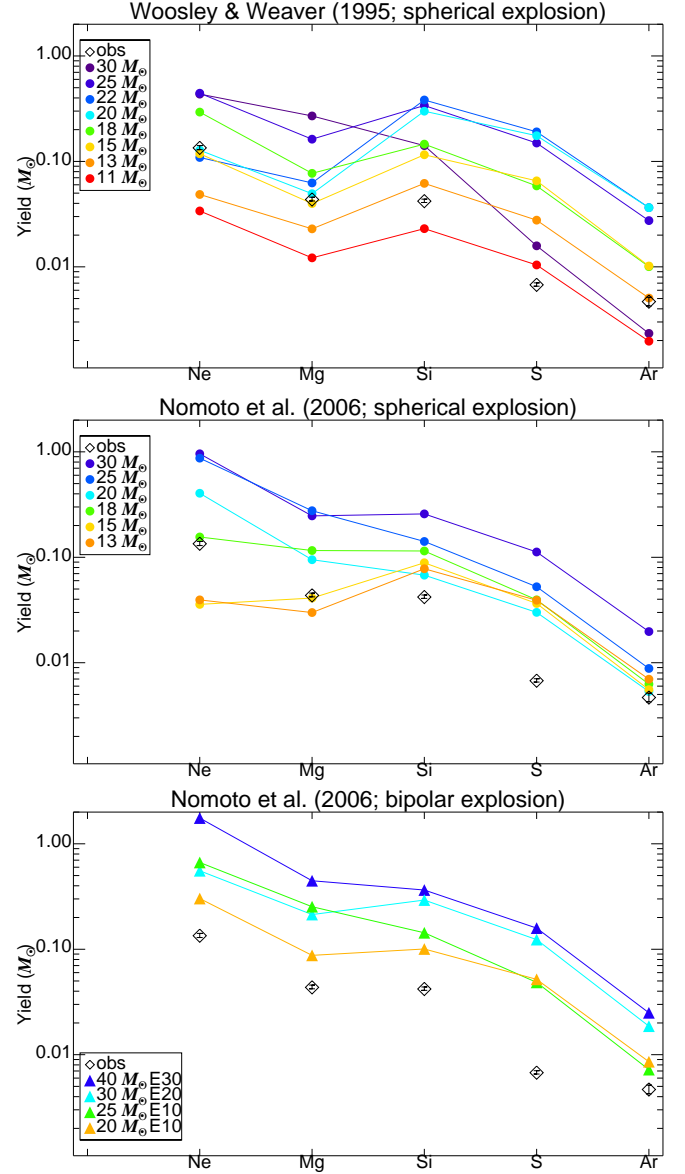}
\caption{
Supernova yields and the masses of the X-ray-emitting elements 
(diamonds with error bars).
The model yields of different progenitor masses (dots for spherical 
explosion and triangle for bipolar explosion) are taken
from the core-collapse nucleosynthesis models of Woosley \& Weaver (1995)
and Nomoto \etal\ (2006).
The bottom panel shows the yields from bipolar models with the kinetic 
energy enhanced by a factor of 10--30 (e.g., E20 stands for $E=20\E{51}$~erg).
}
\label{f:yield}
\end{figure}

\snr\ hosts a CCO, PSR J1852+0040, which is strong evidence of 
core-collapse explosion.
Therefore, its progenitor is a massive ($\gtrsim 8~\Msun$) star born in the MCs.
During its short lifetime ($\sim 10^6$--$10^7$ years in the 
main-sequence stage; Schaller \etal\ 1992),
the massive progenitor launches a strong stellar wind which blows
a stellar wind bubble and sculpts the parent MCs.
The SNR expanding inside the bubble may finally interact with
the molecular cavity wall if the MC has not been dissipated,
and actually several such SNRs have been revealed
during the past several decades (see Chen \etal\ 2014; Zhang \etal\ 2015 
and references therein).
Chen \etal\ (2013) found a linear relationship between
the size of the massive star's molecular cavity and the star's
initial mass ($\Rb$--$M$ relation): 
$p_5^{1/3} \Rb\approx \alpha M-\beta$~pc, where
$\alpha=1.22\pm0.05$, $\beta=9.16\pm1.77$, $p_5\equiv (p/k)/(10^5 \cm^{-3} \K)$, with the mean pressure $p/k$ in the MCs
about $10^5 \cm^{-3} \K$.
This relation can be used to estimate the initial masses
of the progenitors for SNRs interacting with molecular cavity walls
or shells.
\snr\ is associated with the MCs in the velocity range of
95-115~\kms, and the
molecular shell delineates the western radio outer shell
and confines the middle shell.
If the smaller hemisphere corresponding to the bright X-ray/IR
filaments is in contact with the molecular cavity created by the 
progenitor's stellar wind, we can adopt $\Rb\ge R_1\sim 4'$
(8.3~pc; we use $\ge$ here since part of the wind energy may leak into the 
tenuous medium).
The progenitor mass of \snr\ is thus estimated to be $\ge14\pm 2~\Msun$.

Another method to estimate the progenitor mass is to compare
the metal compositions inferred from the X-ray spectra with those 
predicted in the nucleosynthesis models.
This method requires that different metal species are well mixed 
and observed, if the abundances from the global gas properties are 
used for comparison.
According to the \XMMN\ spectral analysis of the global hot
gas, \snr\ has enriched Ne (1.8),  Mg (1.6), Si (1.4) , S (1.7), and 
Ar ($\sim 1.8$).
We first compare the best-fit abundances of Ne, Mg, S, and Ar relative 
to Si (e.g, [Ar/Si]/[Ar/Si]$_\odot$) in \snr\ to those
produced by the progenitors with different masses based on the 
spherical core-collapse nucleosynthesis models of Nomoto \etal\ (2006)
and Woosley \& Weaver (1995).
Differences between the two models are discussed by Kumar \etal\ (2014;
see also Kumar \etal\ 2012).
However, we find that none of the available nucleosynthesis models can 
explain all of the observed metal abundance ratios for this SNR.

We subsequently compare the element masses in the X-ray-emitting gas
with the predicted SN yields in the two nucleosynthesis models, which
can provide at least a lower limit of the SNR's progenitor mass.
The element ($Z$) mass $M_Z$ is estimated from the
abundance $A_Z$ and the mass of the hot component $M_{\rm h}$
($=42\pm 1\Msun$): $M_Z=A_Z M_{\rm h} (M_{Z\odot}/M_\odot)$, 
where $(M_{Z\odot}/M_\odot)$ is the solar mass fraction of the metal $Z$
adopted from Anders \& Grevesse (1989).
As shown in Figure~\ref{f:yield}, a progenitor mass less than $15~\Msun$ 
can be excluded (spherical explosion models), since the mass of 
observed Ne exceeds that predicted in the $\le 15~\Msun$ yield models.
It is consistent with the value $\ge 14\pm 2~\Msun$ given by the $\Rb$--$M$ 
relation (discussed above).
The observed metal masses are no larger than the 20~$\Msun$ yield models
for either the spherical or bipolar explosion (hypernova model with kinetic 
energy 10--30 times of the typical value of $10^{51}$~erg) scenarios.
Under the assumption that most of the ejecta are observed in the X-ray 
band and that the Ne and Mg are well mixed in the SNR, the progenitor 
mass of \snr\ is likely between 15 and 20~$\Msun$.

However, the element masses are not consistent with any group of the 
modeled nucleosynthesis yields.
The first explanation is that the ejecta are not well mixed or asymmetrically 
distributed in the SNR.
Asymmetric distribution of ejecta is present in the SNR as discussed in 
Section~\ref{S:fragment}.
An alternative reason might be that \snr\ is born from an asymmetric 
explosion with normal kinetic energy ($\sim 10^{51}$~erg) and the 
ejecta abundances cannot be explained with current spherical explosion 
or hypernova bipolar explosion model.
Last but not least, the available nucleosynthesis models do not 
consistently provide similar yields, partly because of the assumptions 
made in the calculations. 
Future detailed models incorporating mildly asymmetric explosions
and binary evolution are desirable to provide a more secure 
constraint on the progenitor of \snr\ and other SNRs.

\section{Summary} \label{S:summary}
We have investigated the multi-wavelength emission from the thermal 
composite SNR \snr.
Using the multi-transition CO data covering the whole SNR, we 
study the large-scale molecular environment as well as the small-scale 
structures which cause the asymmetries of the SNR.
We also revisit the 380~ks \XMMN\ MOS data and carry out
imaging and spectroscopic analysis of the X-ray-emitting plasma.
The combined long-exposure X-ray data allow us to study the detailed
distribution of different elements and the hot gas
properties across the SNR.  
The main results are summarized as follows.

\begin{enumerate}
\item 
We provide kinematic and morphological evidence to support the
interaction of SNR \snr\ with the MCs in the velocity range 95-115~\kms: 
(1) broadened \twCO~\Jttt\ line ($\Delta v\approx 12~\km\ps$) is
detected for the first time near the protrusion at the northeastern 
radio boundary;
(2) the morphology agreement between the two \twCO~\Jttt\ filaments
and the radio/IR/X-ray filaments in the east, and between the
western molecular shell and the radio shells;
and (3) the red-shifted \twCO\ lines relative to \thCO, suggesting an 
interaction with the MCs from the foreground SNR.
The molecular gas region (mostly inside a $22'$ region) near \snr\ 
has a mass of $\sim5.7\du^2\times 10^4~\Msun$, which is typical 
for a GMC.

\item 
The overall X-ray-emitting gas can be characterized by a 
cool ($\kTc=0.195^{+0.003}_{-0.002}$~keV) under-ionized plasma 
with $\tauh =6.4\pm0.4 \E{11}~\cm^{-3}\s$ and solar abundances,
plus a hot ($\kTh=0.80\pm 0.01$~keV) plasma with ionization 
timescale of $\tauh=8.1\pm0.1\E{10} \cm^{-3} \s$ and elevated Ne
($1.83^{+0.04}_{-0.05}$), Mg ($1.56\pm 0.03$ ), Si ($1.40\pm 0.02$), 
S ($1.72\pm 0.03$), and Ar ($1.8\pm 0.1$) abundances.
The average densities of the two components are $\sim 2~\cm^{-3}$
and $\sim0.5~\cm^{-3}$, respectively.
The masses are $\sim 280~\Msun$ and $\sim 40~\Msun$, respectively.
Hence, most of the X-ray emission is contributed by the shocked ISM.
\snr\ has a Sedov age of 4.4--6.7~kyr. The mean shock velocity is 
$7.3\E{2}$~\kms.

\item 
The \XMMN\ image of \snr\ reveals many bright X-ray filaments
embedded in a faint halo.
A two-temperature model (generally $\kTc\sim 0.2$~keV and 
$\kTh\sim 0.7$--1.0~keV) is required to describe the small-scale 
regions' spectra all over the SNR.
The filamentary gas has densities  $\sim 4$--$13 \cm^{-3}$ and
$\sim 1$--$2~\cm^{-3}$ for the cool and hot components,
respectively, which are much larger than those of the halo gas
($\sim 1.4$--$2.4~\cm^{-3}$ and $\sim 0.3$--$0.5~\cm^{-3}$
for the cool and hot components, respectively).
The ionization ages of the hot component in the filaments 
(0.5--2.0~kyr) are smaller than in the halo
($\sim 4$--11~kyr).
The X-ray-bright filaments are probably produced by the SNR 
interaction with the dense ambient gas, while the halo forms 
from SNR breaking out to a low-density medium.

\item
The SNR shock propagating into the dense gas produces bright 
filamentary radiation in multiple wavelengths.
The X-ray filaments show good spatial correlation with the 
24~$\um$ IR filaments and part of the radio shells. 
The filamentary mid-IR emission may come from the dust grains
collisionally heated by the hot plasma.

\item Due to shaping by the inhomogeneous environment, \snr\ is 
likely to have a double-hemisphere morphology.
The smaller hemisphere containing bright filaments is at
the back side and projected into the SNR interior.
Projection effect can explain the multiple-shell structures
and the thermal composite morphology of \snr.

\item We find a high-velocity ejecta fragment which shows
distinctly high temperature ($\kTh\sim 1.6^{+0.6}_{-0.3}$~keV) and 
abundances of Ne ($4.4^{+3.5}_{-2.2}$), Mg ($3.6^{+1.7}_{-1.0}$), 
Si ($5.2^{+2.2}_{-1.3}$), S ($5.2^{+1.8}_{-1.1}$), and 
Ar ($4.2^{+2.4}_{-1.8}$).
Its velocity ($1.0^{+0.2}_{-0.1}~\km\ps$) is 32\% larger than 
the average velocity of the blast wave.
The high-velocity ejecta fragment, in addition to the 
asymmetric metal distribution across the remnant, supports
the idea that the SN explosion is intrinsically asymmetric.

\item 
The progenitor mass of \snr\ is estimated to be 
15--$20~\Msun$ by using two methods: (1) the linear relation between 
the progenitor mass and the wind blown bubble size
($\Rb$--$M$ relation), and (2) a comparison between the metal masses
and the yields predicted by nucleosynthesis models.

\end{enumerate}

\begin{acknowledgements}
P.Z. is thankful to Randall Smith for helpful discussions on
the NEI models.
P.Z. and Y.C. thank the support of NSFC grants 11503008, 11590781,
and 11233001 and the 973 Program grant 2015CB857100.
S.S.H. acknowledges support from the Canada Research Chairs program, 
the NSERC Discovery Grants program, and the Canadian Space Agency. 
Z.Y.Z. acknowledges support from the European Research Council (ERC)
in the form of an Advanced Grant, COSMICISM.
Part of this research is conducted during a China Scholarship Council 
award held by P.Z. at the U. of Manitoba. 
This research made use of NASA’s Astrophysics Data System (ADS) 
and the High-Energy Science Archive Research Center (HEASARC) 
maintained at NASA’s Goddard Space Flight Center.

\end{acknowledgements}

\end{document}